\documentclass[10pt,aps,pra,twocolumn,amsmath,amssymb,superscriptaddress]{revtex4-2}

\usepackage{graphicx}
\usepackage{color}
\usepackage{epstopdf}
\usepackage{listings}
\usepackage{braket}
\usepackage{float}
\usepackage{hyperref}
\usepackage{subcaption}
\usepackage{soul}
\usepackage[normalem]{ulem}

\begin{document}
\title{Pitch-controlled reorientational nonlinearity in chiral nematic liquid crystals: a reduced-order model for self-focusing and soliton formation}

\author{Homa Saadatmand}
\thanks{These authors contributed equally.}
\affiliation{Department of Physics, University of Miami, 1320 Campo Sano Ave, Coral Gables, Florida 33146, USA}
\author{M. Javad Zakeri}
\thanks{These authors contributed equally.}
\affiliation{The College of Optics and Photonics, University of Central Florida, Orlando, Florida 32816, USA}
\author{Ahmed Sameh Ahmed}
\affiliation{Department of Electrical and Computer Engineering, University of Miami, 1320 Campo Sano Ave, Coral Gables, Florida 33146, USA}
\author{Suraj Bhandari}
\affiliation{Department of Physics, University of Miami, 1320 Campo Sano Ave, Coral Gables, Florida 33146, USA}
\author{Loubna Benkoula}
\affiliation{Department of Physics, University of Miami, 1320 Campo Sano Ave, Coral Gables, Florida 33146, USA}
\author{Ameer B. Batarseh}
\affiliation{The College of Optics and Photonics, University of Central Florida, Orlando, Florida 32816, USA}
\author{Andrea Blanco-Redondo}
\affiliation{The College of Optics and Photonics, University of Central Florida, Orlando, Florida 32816, USA}
\author{Miroslaw Karpierz}
\affiliation{Faculty of Physics, Warsaw University of Technology, Koszykowa 75, 00-662 Warsaw, Poland}
\author{Pawel S. Jung}
\email{pjung@miami.edu}
\affiliation{Department of Physics, University of Miami, 1320 Campo Sano Ave, Coral Gables, Florida 33146, USA}
\affiliation{Faculty of Physics, Warsaw University of Technology, Koszykowa 75, 00-662 Warsaw, Poland}

\date{March 9, 2026}

\begin{abstract}
We present a reduced-order semi-analytical model for reorientational nonlinearity in chiral nematic liquid crystals, showing that the chiral pitch acts as the dominant physical length scale governing the onset of nonlinear self-focusing and soliton formation. Starting from the full Frank–Oseen equation, we derive a closed-form expression for the optically induced molecular rotation that captures the essential saturable response of the medium while reducing computational cost by more than two orders of magnitude compared with standard relaxation-method solvers. Despite its simplicity, the model reproduces the essential features of the numerically obtained nonlinear refractive index, the onset of self-localization, and the transition from discrete to continuous solitons in one and two dimensions. It further predicts the formation of fully localized astigmatic nematicons with only minor shifts in the self-localization threshold due to the neglect of nonlocal effects. The proposed model provides direct physical insight into light–matter interactions with soft matter media and offers a computationally efficient tool for the design and optimization of nonlinear photonic devices.

\end{abstract}

\begin{center}
\footnotesize
\copyright~2026 American Physical Society. 
This is the accepted manuscript of the following article:

H. Saadatmand, M. J. Zakeri, A. S. Ahmed, \textit{et al.},
``Pitch-controlled reorientational nonlinearity in chiral nematic liquid crystals:
A reduced-order model for self-focusing and soliton formation,''
Phys. Rev. A \textbf{113}, 033511 (2026).

The final published version is available at
\href{https://doi.org/10.1103/lsbq-lyrt}
{https://doi.org/10.1103/lsbq-lyrt}.
This manuscript version is posted with permission for non-commercial scholarly use.
\end{center}

\maketitle


\section{Introduction}
Beam self-focusing~\cite{akhmanov1968self} is a fundamental phenomenon observed in nonlinear optics in which the refractive index of a medium increases in regions of higher light intensity. Similar self-focusing and self-trapping effects are also observed in plasmas~\cite{Litvak1965, Shukla1992}, Bose–Einstein condensates~\cite{Dalfovo1999, Pitaevskii2003}, acoustic systems~\cite{Achilleos2015}, and hydrodynamic waves~\cite{kuznetsov1986}, highlighting the universality of nonlinear self-localization phenomena~\cite{Shukla1984, Kivshar2003, Abdullaev2004, Wabnitz2017}. This effect counteracts diffraction, enabling beam self-localization and the formation of spatial solitons in nonlinear media such as fused silica, chalcogenide glasses, photorefractive crystals, semiconductors, and liquid crystals, to name a few~\cite{sukhorukov2002nonlinear}. In conventional Kerr-type materials, where the change in refractive index is proportional to the optical intensity ($\Delta n \propto I$), self-focusing can lead to catastrophic beam collapse once the critical power is exceeded~\cite{bang1999collapse}. Such collapse not only hinders stable soliton propagation but can also induce permanent material damage. Nonlinear media with saturable and/or nonlocal responses provide a route to prevent collapse~\cite{bang2002collapse}: in these systems, the refractive-index change either saturates at high intensities, depends on the spatially averaged optical field, or both—thereby arresting collapse and allowing the stable propagation of self-trapped beams. The general mechanism of collapse arrest in nonlocal nonlinear media was first formalized through the concept of accessible solitons by Snyder and Mitchell~\cite{Snyder1997}, later expanded by Kr{\'o}likowski et al.~\cite{Krolikowski2001} and experimentally verified by Conti, Peccianti, and Assanto~\cite{conti2003route}.

\begin{figure}[!ht]
\centering
\includegraphics[width=0.46\textwidth]{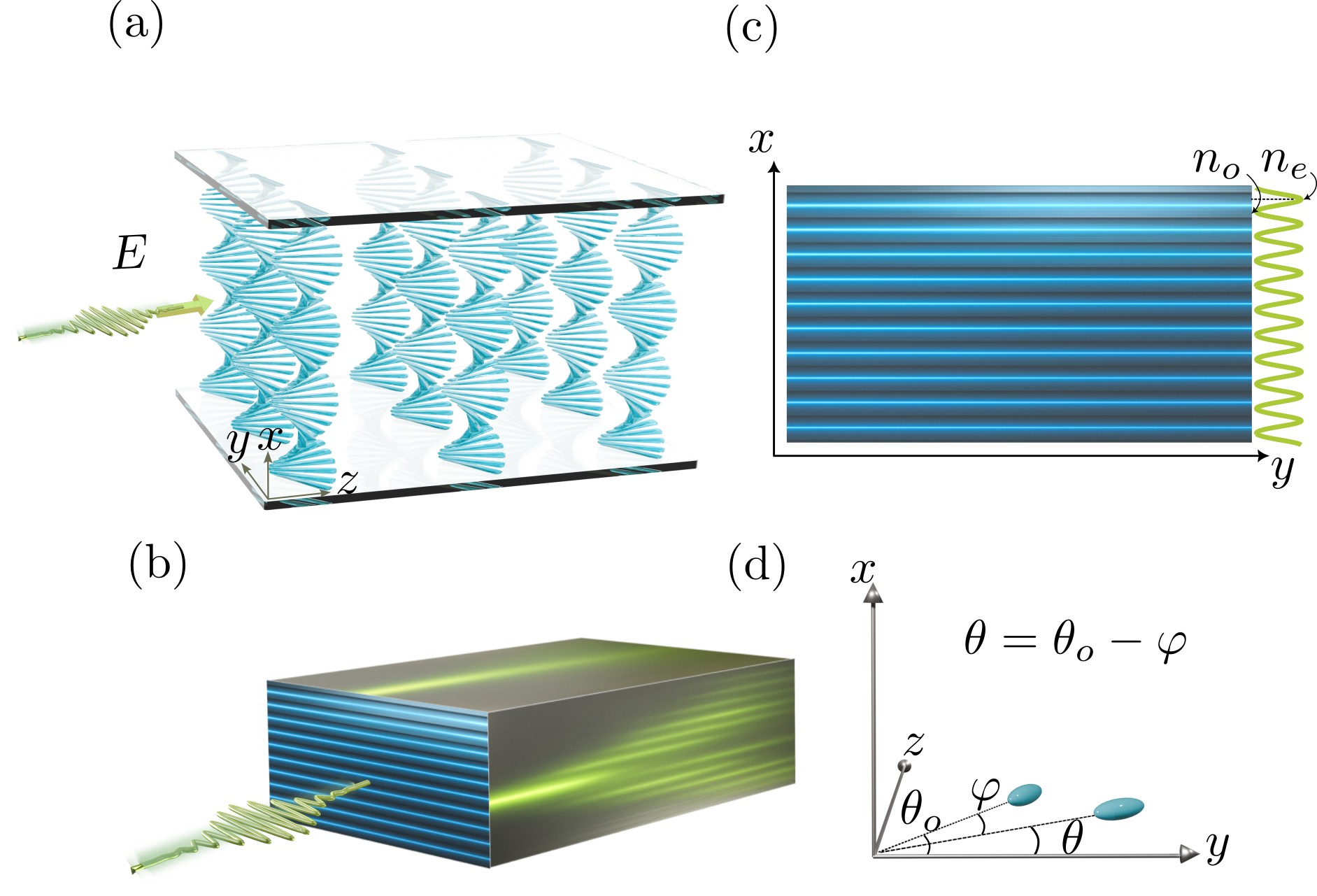}
\caption{
(a) Three-dimensional schematic of the chiral nematic liquid crystal (ChNLC) cell, where molecules form a periodic helical structure along the \textit{z}-axis, leading to a refractive index modulation along the \textit{x}-direction. (b) Beam propagation through the ChNLC medium, where the optical field experiences discrete diffraction along \textit{x} and continuous diffraction along \textit{y}, producing anisotropic spreading. (c) Top view illustrating the effective periodic index modulation along \textit{x} (resulting in discrete diffraction), while the medium remains uniform along \textit{y}. (d) Molecular reorientation geometry in the \textit{yz}-plane, where the director angle is defined as $\theta = \theta_0 - \varphi$, linking the initial alignment $\theta_0$ to the optically induced rotation $\varphi$.}
\label{Figure1}
\end{figure}

Nematic liquid crystals (NLCs) are exemplary nonlinear optical media~\cite{peccianti2006nonlinear,assanto2009nematicons,peccianti2012nematicons,braun1993}. Their nonlinear response is simultaneously strong, saturable, and nonlocal~\cite{conti2003route,conti2004observation}, enabling the formation of bright, bell-shaped spatial solitons~\cite{piccardi2024nematonics,hutsebaut2004single,laudyn2015nonlinear,warenghem2008thermo}, as well as more complex self-trapped structures such as multipeak solitons~\cite{jung2017supermode,batarseh2025crossover,jung2024giant}—all at milliwatt-level powers. This unique combination suppresses catastrophic collapse and makes NLCs an ideal platform for studying nonlinear light–matter interactions. Foundational descriptions of nematicon dynamics and nonlocal reorientational effects can be found in Alberucci and Assanto’s analytical treatments~\cite{Alberucci2009} and the classical nematic theory of de Gennes and Prost~\cite{DeGennes1993}. Within this class, chiral nematic liquid crystals (ChNLCs) are particularly attractive: their periodic helical molecular structure induces discrete diffraction along one transverse axis while maintaining continuous diffraction along the other, thereby enabling the formation of quasi-astigmatic solitons with hybrid spatial localization~\cite{oswald2005nematic,laudyn2009nematicons,laudyn2014power,laudyn2016quasi}.

Despite their promise, existing models describing reorientational-nonlinearity-induced self-focusing in ChNLCs are computationally demanding, often requiring full numerical solutions of the reorientation equations. This complexity can obscure the underlying physics, slow parameter exploration, and hinder the design of liquid-crystal-based photonic devices~\cite{ma2022self,kang2024liquid,zhang2023advanced,yang20213d,perumbilavil2019spatiospectral}.

In this work, we develop a compact reduced-order, semi-analytical model that captures the essential physics of saturable reorientational self-focusing in chiral nematic liquid crystals while significantly reducing computational cost. We show that the chiral pitch governs the effective nonlinear confinement by setting the characteristic length scale for self-localization, enabling the introduction of an effective chiral width $\omega_{\mathrm{ch}}$. This reduced description provides direct physical insight into the coupling between optical fields and molecular reorientation, allows rapid exploration of parameter space, and offers an efficient framework for understanding self-focusing and soliton formation in pitch-modulated nematic media.

\section{Linear Light Propagation in a Chiral Nematic Liquid Crystal}

We begin by considering the propagation of a paraxial, extraordinarily polarized optical wave packet in the forward $z$-direction through a ChNLC cell, as illustrated in Fig.~\ref{Figure1}(a). The cell consists of two parallel plates separated by a distance $d$, with boundary conditions that fix the initial molecular alignment. For a chiral nematic crystal, the initial director orientation is given by $\theta_0(x, y) = \frac{2\pi}{\Lambda} x$, where $\Lambda$ is the pitch, i.e., the distance over which the director undergoes a full $360^\circ$ rotation. In this configuration, the refractive index is periodic along the $x$-axis and uniform along the $y$-axis, as shown in Fig.~\ref{Figure1}(b), and the resulting refractive index distribution can be expressed as

\begin{equation}
    n(x,y)=\frac{n_o n_e}{\sqrt{n_e^2\sin^2\theta(x,y)+n_o^2\cos^2\theta(x,y)}}, \label{nnn}
\end{equation}
where $n_o$ and $n_e$ denote the ordinary and extraordinary refractive indices of the ChNLC, respectively.

Under these conditions, and neglecting the walk-off effect due to small liquid-crystal anisotropy, the light dynamics of an extraordinarily polarized beam can be described by the paraxial wave equation~\cite{assanto2003spatial}:
\begin{equation}\label{paraxial}
    i k_0 n_o \frac{\partial E}{\partial z} 
    + \frac{\partial^2 E}{\partial x^2} 
    + \frac{\partial^2 E}{\partial y^2} 
    + k_0^2 \big(n^2 - n_o^2\big) E = 0,
\end{equation}
where $k_0 = 2\pi / \lambda$ is the free-space wavenumber, and $E(x, y, z)$ is the slowly varying envelope of the electric field.

As shown schematically in Fig.~\ref{Figure1}(b)--(c), under linear (low-power) conditions, a $y$-polarized beam launched along $z$ experiences a refractive index that varies periodically across $x$, ranging from $n_o$ to $n_e$. This $\Lambda/2$-periodic modulation effectively forms an array of planar, one-dimensional, graded-index waveguides. A beam launched into a single waveguide may remain laterally confined or—under weak guiding conditions—propagate via weak coupling between neighboring waveguides along $x$. Similar coupling behavior was first demonstrated in optical waveguide arrays by Christodoulides and Silberberg~\cite{christodoulides2003discretizing}, and later realized in nematic liquid-crystal arrays by Karpierz and Assanto~\cite{fratalocchi2004discrete}. In our configuration, such coupling gives rise to discrete diffraction in the $x$-direction, while continuous diffraction occurs in the homogeneous $yz$-plane, as illustrated in Fig.~\ref{Figure1}(b)--(c). Similar discrete diffraction effects and photonic-band coupling in helical or chiral nematic systems have been analyzed by Bregar~\cite{Bregar2018} 
and Kopp \textit{et~al.}~\cite{Kopp1998}, establishing a direct link between pitch geometry and spatial light dynamics.

\section{Nonlinear Light Propagation in a ChNLC}

In the nonlinear regime, the electric field of a $y$-polarized beam, oscillating in the $yz$-plane with phase fronts orthogonal to the propagation vector $k$, induces molecular dipoles. In a positive uniaxial nematic liquid crystal (NLC), the resulting optical torque reorients the molecules toward the field vector, changing their orientation to $\theta = \theta_0 - \varphi$, where $\theta_0$ is the initial director orientation and $\varphi$ is the optically induced change in orientation, as shown in Fig.~\ref{Figure1}(d). This reorientation increases the effective refractive index experienced by the beam, producing a strong self-focusing response. Such light-driven molecular reorientation can lead to self-localization of the beam into spatial solitons (nematicons) in both uniform and chiral NLC systems.

To model this reorientational nonlinearity, we assume a single elastic constant $K$, since molecular reorientation for a $y$-polarized beam occurs predominantly in the $yz$-plane. By minimizing the Frank-Oseen free energy density via the Euler–Lagrange equation, the steady-state molecular reorientation is described by~\cite{sala2012modeling}
 
\begin{equation}
    \nabla^2 \theta - \frac{\Delta \varepsilon \varepsilon_0}{2K} f(\vec{E}) = 0, \label{Eq1}
\end{equation}
where $\varepsilon_0$ is the vacuum permittivity, $\Delta\varepsilon = n_e^2 - n_o^2$, and
\begin{equation}
   f(\vec{E}) = 2E_x E_z \cos 2\theta + \sin 2\theta \big(|E_y|^2 - |E_z|^2 \big).
\end{equation}
 For a paraxial, $y$-polarized beam propagation (where $E_x = E_z = 0$ and $E_y = E$) with slowly varying intensity along $z$, Eq.~\eqref{Eq1} reduces to

\begin{equation}
    \frac{\partial^2 \theta}{\partial y^2} + 
    \frac{\partial^2 \theta}{\partial x^2} - 
    \frac{\Delta \varepsilon \varepsilon_0}{2K} |E|^2 \sin 2\theta = 0. \label{Eq2}
\end{equation}
Under continuous-wave excitation, the director dynamics can be treated in a steady-state (adiabatic) approximation, as the molecular reorientation time scale is slow compared with optical propagation but fast relative to changes in the beam envelope. Solving Eq.~\eqref{Eq2} under the relevant boundary conditions yields the equilibrium director profile, which determines the nonlinear refractive index distribution experienced by the beam. In practice, this requires numerical iterative methods—such as the relaxation method—which are accurate but computationally intensive to implement.

\section{Reduced-order model for pitch-controlled reorientational nonlinearity}

Building on the reorientation equation derived in Eq.~\eqref{Eq2}, we now seek a semi-analytical reduced-order model that avoids the computational overhead of iterative numerical solvers. Guided by numerical solutions of Eq.~\eqref{Eq2}, we propose a phenomenological expression for the optically induced molecular reorientation $\varphi(x, y)$ (where $\theta = \theta_0 - \varphi$) that captures both the slowly varying envelope and the rapid helical modulation intrinsic to the ChNLC structure (Fig.~\ref{fig:function}). Motivated by the numerical solutions, we propose the following ansatz:

\begin{equation}\label{phi_function}
    \varphi(x,y) = \varphi_0 
        \cos\!\!\left(\frac{\pi x'}{2\omega_x}\right)
        \sin\!\!\left(\frac{\pi x'}{2\omega_{\mathrm{ch}}}\right)
        \cos\!\!\left(\frac{\pi y'}{2\omega_y}\right),
\end{equation}
where $\varphi_0$ is the maximum reorientation angle, $\omega_x$ and $\omega_y$ denote the characteristic widths of the reorientation region along $x$ and $y$, respectively, and $(x_0, y_0)$ defines the beam center, such that $x' = x - x_0$ and $y' = y - y_0$. The first and third terms describe the broad spatial envelope of the molecular reorientation, while the middle term captures the high-frequency chiral modulation at the pitch scale $\omega_{\mathrm{ch}}$.

Substituting Eq.~\eqref{phi_function} into Eq.~\eqref{Eq2} yields

\begin{equation}\label{Eq4}
    \frac{2\varphi}{\cos 2\varphi}\left( I_{\mathrm{sat}} 
    + B \right) 
    + \tan(2\varphi)\cos 2\theta_0 |E|^2 
    = \sin 2\theta_0 |E|^2, 
\end{equation}
where $I_{\mathrm{sat}} = 
\frac{\pi^2 K}{4 \varepsilon_0 \Delta \varepsilon}
\!\left(
\frac{1}{\omega_x^2} + \frac{1}{\omega_y^2} + \frac{1}{\omega_{\mathrm{ch}}^2}
\right) $, and $B = 
\frac{K \pi^{2}}{2\varepsilon_0 \Delta \varepsilon} 
\frac{1}{\omega_{\mathrm{ch}} \omega_x}
\tan\!\!\left(\frac{\pi x'}{2 \omega_x}\right)
\cot\!\!\left(\frac{\pi x'}{2 \omega_{\mathrm{ch}}}\right).$
Close to the beam center, where $B \ll I_{\mathrm{sat}}$, the $B$-term may be neglected without loss of accuracy. For small $\varphi$, we then approximate $2\varphi \approx \sin 2\varphi$, which leads to a compact, closed-form solution for the reorientation angle:
\begin{equation}\label{phi_equation1}
      \varphi = \frac{1}{2}\tan^{-1}\!\!\left(\frac{\sin 2\theta_0 I}{\frac{\pi}{2} + \cos 2\theta_0 I}\right), 
\end{equation}
where $I=\frac{|E|^2}{I_{\mathrm{sat}}}$. Moreover, for small birefringence $n_e^2-n_o^2$, Eq.~\eqref{nnn} can be approximated as $n=n_o+(n_e-n_o)\cos^2 \theta = n_o+0.5(n_e-n_o)(1+\cos(2\theta_0-2\varphi) )$~\cite{jung2017supermode}, which leads to the following expression (The full derivation is provided in the Appendix).
\begin{equation}\label{napprox}
    n \approx n_o + \frac{1}{2}(n_e - n_o)\left[1 - \frac{\frac{\pi}{2} \cos 2\theta_0 + I}{\sqrt{\left(\frac{\pi}{2}\right)^2 + \pi I \cos 2\theta_0 + I^2}} \right].
\end{equation}

This analytical approximation reproduces the numerically obtained director reorientation and nonlinear refractive index change with high accuracy while being substantially more computationally efficient, as demonstrated in the following sections. It therefore serves as a practical tool for exploring parameter space and guiding experimental design.

\subsection{Results}
To evaluate the accuracy and applicability of Eqs.~\eqref{Eq4}--\eqref{napprox}, we first estimate the unknown characteristic width parameters $\omega_x$, $\omega_y$, and $\omega_{\mathrm{ch}}$. To this end, we performed a series of numerical simulations of Eq.~\eqref{Eq4} for chiral nematic liquid crystals (ChNLCs) across a broad range of physical parameters. The simulations explored pitches between $4~\mu\mathrm{m}$ and $20~\mu\mathrm{m}$, birefringence values spanning $0.02 < n_e - n_o < 0.4$, and input Gaussian beam waists in the range $1~\mu\mathrm{m} < \omega_0 < 50~\mu\mathrm{m}$, with optical powers ranging from the microwatt to the hundreds of milliwatt regime.

\begin{figure}[!ht]
\centering
\includegraphics[width=0.45\textwidth]{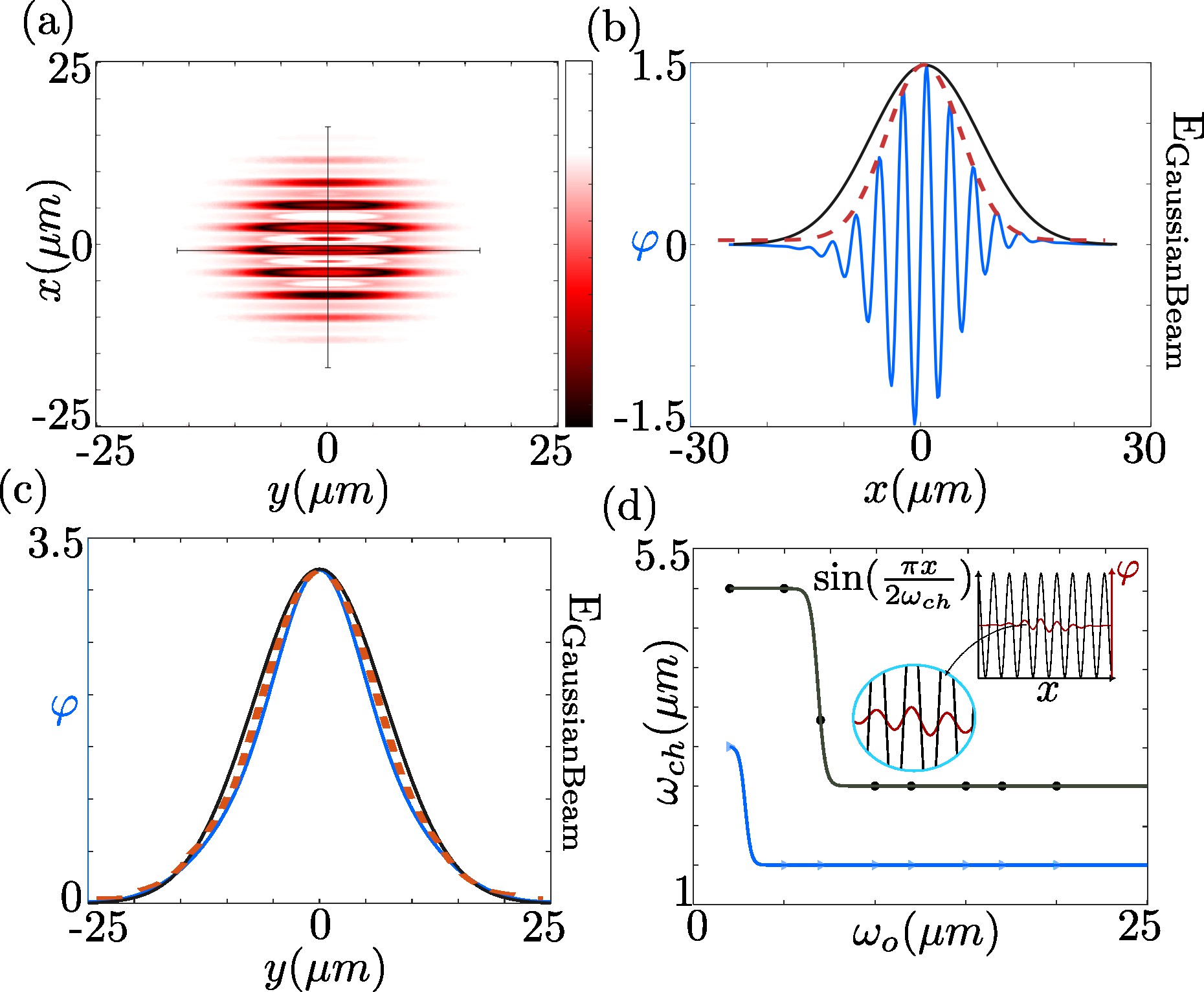}
\caption{
(a) Change in the molecular orientation angle $\varphi(x,y)$ in the transverse plane induced by a Gaussian beam, calculated from Eq.~\eqref{Eq2}.
(b)--(c) The corresponding cross-sections of the reorientation angle $\varphi$ along
(b) the $y$-axis for $x = 0$ and
(c) the $x$-axis for $y = 0$.
The blue solid curve shows the numerically obtained $\varphi$, the red dashed curve represents the fitted envelope, and the black solid curve corresponds to the Gaussian beam electric-field profile.
(d) Extracted $\omega_{\mathrm{ch}}$ as a function of the input Gaussian beam waist $\omega_0$ (symbols), together with the empirical fit from Eq.~(10) (solid curves) for $\Lambda = 6\,\mu\mathrm{m}$ (blue) and $\Lambda = 16\,\mu\mathrm{m}$ (black). Inset: comparison of the simulated reorientation profile $\varphi(x)$ (red) with $\sin\!\left(\pi x / 2\omega_{\mathrm{ch}}\right)$ (black), illustrating how $\omega_{\mathrm{ch}}$ is obtained from the spatial period of $\varphi(x)$.
}

\label{fig:function}
\end{figure}

The results show that the characteristic widths $\omega_x$ and $\omega_y$ of the molecular reorientation profile closely follow the input beam waist $\omega_0$, confirming that the nonlinear response follows the transverse intensity envelope. In contrast, the chiral modulation length $\omega_{\mathrm{ch}}$ exhibits a more complex dependence on $\omega_0$, displaying a distinct crossover at a critical beam waist. This crossover occurs when the optical field becomes confined within a single waveguide layer of width $\Lambda/2$, as illustrated in Fig.~\ref{fig:function}(d). To quantitatively describe this behavior, we introduce a unified semiempirical expression for the effective pitch-to-reorientation scaling,
\begin{equation}
    \omega_{\mathrm{ch}}=
    \frac{\Lambda}{\Lambda_6}\left(\alpha + \frac{\beta}{1 + \exp\!\big[\frac{\Lambda_6-\sigma}{\Lambda-\sigma}\zeta \omega_{0}- \zeta \eta\big]}\right), 
    \label{wch_function}
\end{equation}
where $\alpha$, $\beta$, $\zeta$, $\sigma$, and $\eta$ are fitting parameters obtained from simulation data, with the following values:
$\alpha = 1.51~\mu\mathrm{m}$,
$\beta = 1.50~\mu\mathrm{m}$,
$\zeta = 6.20~\mu\mathrm{m^{-1}}$,
$\sigma = 2.5~\mu\mathrm{m}$,
$\eta = 2.85~\mu\mathrm{m}$,
and $\Lambda_6 = 6~\mu\mathrm{m}$.
Here, $\Lambda$ represents the pitch under consideration.

The parameter $\omega_{\mathrm{ch}}$ represents an effective chiral confinement length rather than a microscopic material constant. Physically, it reflects whether the optically induced reorientation is averaged over multiple helical layers or becomes confined within a single half-pitch channel. The resulting crossover is governed by the geometric interplay between the beam width and the chiral pitch, which motivates the phenomenological logistic form in Eq.~\eqref{wch_function}. We emphasize that this form is not universal, but provides an effective description of the pitch-controlled confinement transition within the parameter ranges considered here. Alternative smooth crossover functions yield similar qualitative behavior, but the logistic form offers a compact and convenient parametrization of this transition.

Within this range, the proposed relation captures the observed dependence of the reorientation spatial frequency $\omega_{\mathrm{ch}}$ on the input beam width $\omega_0$ for pitches in the range $4~\mu\mathrm{m} \leq \Lambda \leq 20~\mu\mathrm{m}$. Fig.~\ref{fig:function}(d) compares the simulated reorientation profile $\varphi(x)$ (red curve) with the corresponding spatial period of $\sin\!\big(\pi x / 2\omega_{\mathrm{ch}}\big)$ (black curve), confirming that the reduced description captures the essential physics of the reorientation process.

\subsubsection{Comparison of Bright Soliton Solutions Using Exact and Simplified Models}

To validate the accuracy of our analytical approximation, we compared bright soliton solutions obtained from the full numerical model with those predicted by the simplified formulation of Eqs.~\eqref{Eq4}--\eqref{napprox}. Soliton profiles were computed using the imaginary-time finite-difference beam propagation method (iFDBPM)~\cite{jungling1994study}, implemented by substituting $z \rightarrow i z$ in the paraxial equation Eq.~\eqref{paraxial}. At each iteration, the electric field was normalized to the desired soliton power, and propagation was continued until the field converged to a stationary soliton solution.
\begin{figure}[htbp]
\centering
\includegraphics[width=0.48\textwidth]{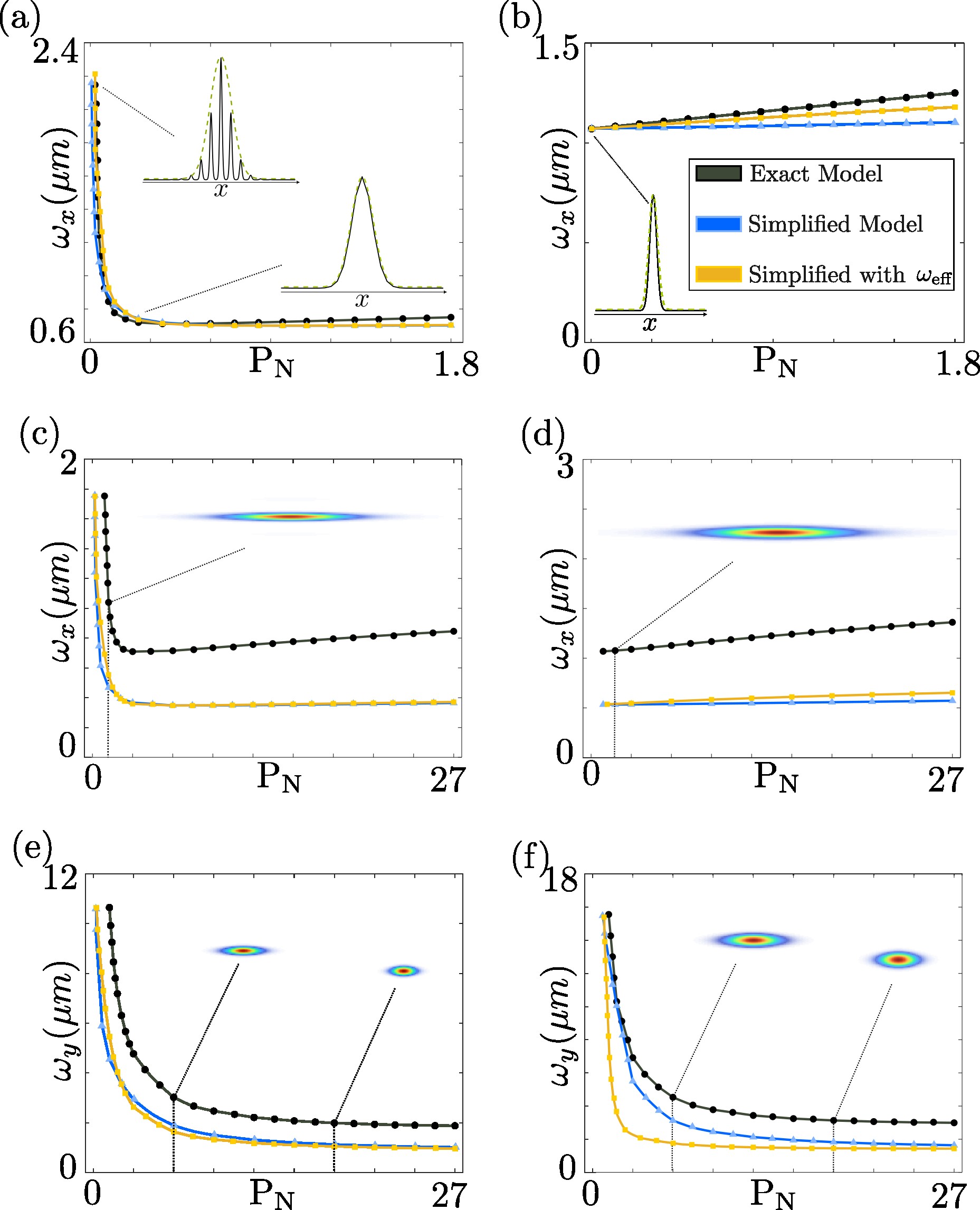}
\caption{
Comparison of bright soliton profiles obtained from the full numerical model (black curves), the reduced-order model (blue curves), and the effective-width approximation (yellow curves) for different chiral pitch values and transverse dimensionalities. The soliton power is normalized as $\mathrm{P}_\mathrm{N} = \gamma \frac{\varepsilon_0 \Delta \varepsilon}{2K} \iint |E|^2 dr = \gamma \frac{\varepsilon_0 \Delta \varepsilon}{2K} \mathrm{P}$, where $\gamma = 1~\mu\mathrm{m}$ and $dr = dx$ for 1D, and $\gamma = 1$ with $dr = dx\,dy$ for 2D. Panels (a) and (b) show one-dimensional soliton profiles, while panels (c)--(f) display the corresponding two-dimensional intensity distributions at different soliton powers. Results are shown for two different pitches: $\Lambda = 6~\mu\mathrm{m}$ in panels (a), (c), and (e), and  $\Lambda = 16~\mu\mathrm{m}$ in panels (b), (d), and (f). The reduced-order description reproduces the essential qualitative features of the soliton width and shape, while quantitative differences become more pronounced for larger pitch values and in the fully two-dimensional regime, reflecting the approximations inherent in the reduced model.}
\label{fig:Figure_1D}
\end{figure}
For the exact model, the corresponding equilibrium director profile was obtained by numerically solving Eq.~\eqref{Eq2} using a relaxation method. In the simplified model, the nonlinear index change was calculated directly from the closed-form expression for the reorientation angle $\varphi(x, y)$ in Eq.~\eqref{phi_equation1}. At each propagation step, the parameters $\omega_x$ and $\omega_y$ were determined from the second moments of the intensity distribution, while $\omega_{\mathrm{ch}}$ was obtained from Eq.~\eqref{wch_function} with the substitution $\omega_0 = \omega_x$. 

Moreover, for additional simplification, we propose a single effective width parameter $\omega_{\mathrm{eff}} = \Lambda / 8$, such that
\begin{equation}
    \frac{1}{\omega_{\mathrm{eff}}^{2}} \approx 
    \frac{1}{\omega_x^{2}} + \frac{1}{\omega_y^{2}} + \frac{1}{\omega_{\mathrm{ch}}^{2}},    
\end{equation}
providing a compact approximation that eliminates the need to compute each characteristic width separately, while preserving good agreement with the full numerical solution.

\begin{figure*}[htbp]
    \centering
    \includegraphics[width=1\textwidth]{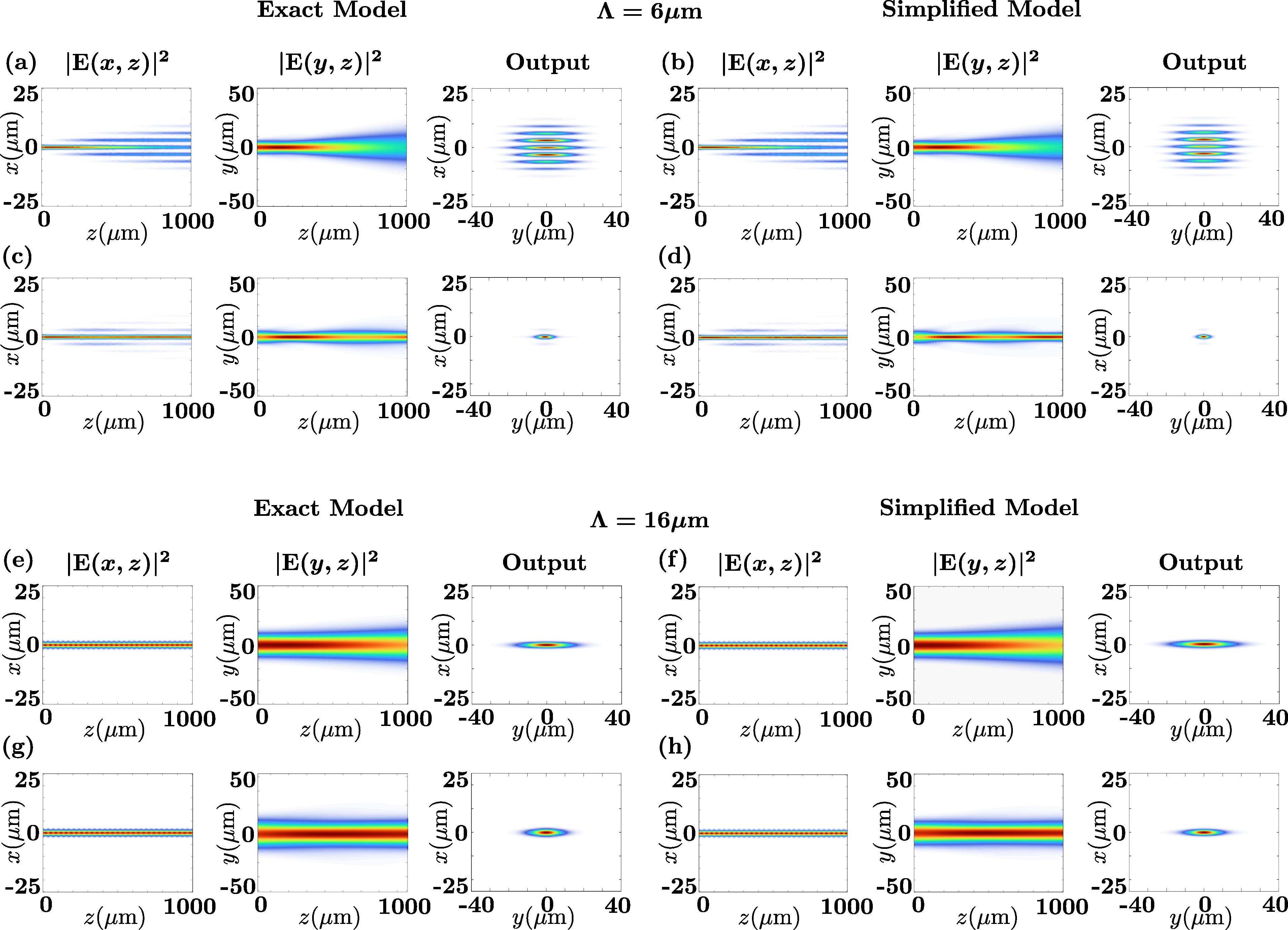}
\caption{
Comparison of nonlinear light dynamics obtained using the exact numerical model (left) and the reduced-order model (right), employing the approximation $\omega_{\mathrm{eff}} = \Lambda / 8$.
For each case, the panels show the beam evolution along $z$ in the $xz$ plane at $y = 0$ (left), the $yz$ plane at $x = 0$ (middle), and the transverse output intensity profile at $z = 1~\mathrm{mm}$ (right).
In all simulations, the launched beam was elliptical, with parameters:
for $\Lambda = 6~\mu\mathrm{m}$, $\omega_{0x} = 1~\mu\mathrm{m}$ and $\omega_{0y} = 6~\mu\mathrm{m}$ (a)--(d);
and for $\Lambda = 16~\mu\mathrm{m}$, $\omega_{0x} = 1.5~\mu\mathrm{m}$ and $\omega_{0y} = 10~\mu\mathrm{m}$ (e)--(h).
The normalized input powers were, for pitch $\Lambda = 6~\mu\mathrm{m}$: (a)~3.0, (b)~1.5, (c)~4.6, (d)~2.8;
and for pitch $\Lambda = 16~\mu\mathrm{m}$: (e)~1.2, (f)~1.5, (g)~2.5, (h)~3.7. 
Comparisons are made at comparable propagation regimes rather than identical normalized powers; the observed differences in required power reflect a systematic shift in the self-localization threshold arising from the neglect of full spatial nonlocality in the reduced-order approximation.
These deviations become more pronounced at larger pitch values, where elastic nonlocality plays a stronger role in the reorientational response.
}
\label{light-dynamics}
\end{figure*}

To benchmark our approach, we considered the nematic liquid crystal 1110~\cite{dabrowski1110mixture} with $n_o = 1.44$, $n_e = 1.49$, and twist elastic constant $K = 8~\mathrm{pN}$. We investigated two distinct regimes:  
(i) the strong-coupling regime, corresponding to a small pitch, in which light is strongly coupled between adjacent waveguides, resulting in pronounced discrete diffraction in the linear propagation regime; and  
(ii) the weak-coupling regime, corresponding to a large pitch, in which coupling between adjacent channels becomes negligible, and the beam remains confined to a single waveguide over a propagation distance of approximately $1~\mathrm{mm}$ at $\lambda = 532~\mathrm{nm}$.  

Studying both regimes allows us to validate the proposed model across the full range of coupling strengths, ensuring that it captures the essential physics of soliton behavior--from multihump discrete solitons to single-peak solitons--in the $xz$-plane.

Having established the relevant material parameters and coupling regimes, we next compare the bright soliton solutions obtained from the full numerical model with those predicted by the simplified analytical approach, as shown in Fig.~\ref{fig:Figure_1D}. Our analysis of the soliton envelope width as a function of normalized soliton power demonstrates that the simplified model remains highly accurate and computationally efficient, even when employing the effective-width approximation $\omega_{\mathrm{eff}}$. 
This behavior is confirmed for the one-dimensional case—where the $y$-dimension is neglected for simplicity, as illustrated in Fig.~\ref{fig:Figure_1D}(a) and (b). In Fig.~\ref{fig:Figure_1D}(a), with a small pitch ($\Lambda = 6~\mu\mathrm{m}$), strong optical coupling between adjacent waveguides leads to multihump solutions at low power. As the power increases, the envelope width decreases, and the field evolves into a single-hump soliton. In contrast, for a large pitch ($\Lambda = 16~\mu\mathrm{m}$), Fig.~\ref{fig:Figure_1D}(b) shows that the soliton is already well confined within a single waveguide even at low power. Further increases in power cause a slight broadening of the soliton due to nonlinear index saturation, which effectively widens the waveguide and results in a gradual increase in soliton width.

These observations confirm that the simplified model reproduces the qualitative transition from multi-hump discrete solitons to single-channel solitons across the full range of input powers and coupling strengths in the one-dimensional case. After demonstrating the reliability of the model in one dimension, we now extend our analysis to the two-dimensional (2D) case, where the light beam launched into the NLC cell during evolution experiences both discrete diffraction along $x$ and continuous diffraction along $y$, as shown in Fig.~\ref{Figure1}(b)~\cite{laudyn2016quasi}. This more stringent scenario allows us to test whether the simplified model can accurately predict the formation of astigmatic solitons, as previously observed experimentally~\cite{laudyn2016quasi,laudyn2014power}. In doing so, Fig.~\ref{fig:Figure_1D}(c)--(f) shows the 2D soliton intensity distributions obtained from both the exact numerical model and the simplified approaches. The comparison reveals good agreement in both beam widths ($\omega_x$ and $\omega_y$) corresponding to astigmatic shapes, demonstrating that the simplified model captures the combined effects of discrete diffraction along $x$ and continuous diffraction along $y$, and reliably predicts the formation of fully localized astigmatic nematicons under realistic experimental conditions.

\subsubsection{Light Dynamics}

With the stationary soliton solutions validated, we now examine the full nonlinear beam dynamics predicted by the model. This analysis allows us to assess whether the reduced-order approach accurately reproduces the power-dependent transition from discrete diffraction to robust self-localization. To this end, we investigated the nonlinear light dynamics in the two-dimensional (2D) 1110 ChNLC structure for both pitch values ($\Lambda = 6~\mu\mathrm{m}$ and $16~\mu\mathrm{m}$), using both the exact numerical model and the reduced-order model, as shown in Fig.~\ref{light-dynamics}.

In the simulations, a Gaussian beam was launched into the central waveguide (at $x = 0, y = 0$), and its evolution along the $z$ direction was examined over a range of input powers. In the low-power (quasi-linear) regime and for small pitch values ($\Lambda = 6~\mu\mathrm{m}$), the light coupling coefficient between neighboring waveguides along the $x$-direction is strong, resulting in pronounced discrete diffraction along $x$. In contrast, due to the absence of periodic modulation along $y$, the beam undergoes continuous diffraction in that direction. Both effects lead to significant transverse spreading and multiple intensity peaks along $x$ in the output cross-section ($x$--$y$ plane) at $z = 1~\mathrm{mm}$, as shown in Fig.~\ref{light-dynamics}(a) for the exact model and Fig.~\ref{light-dynamics}(b) for the corresponding simplified model.
\begin{figure}[!ht]
\centering
\includegraphics[width=0.4\textwidth]{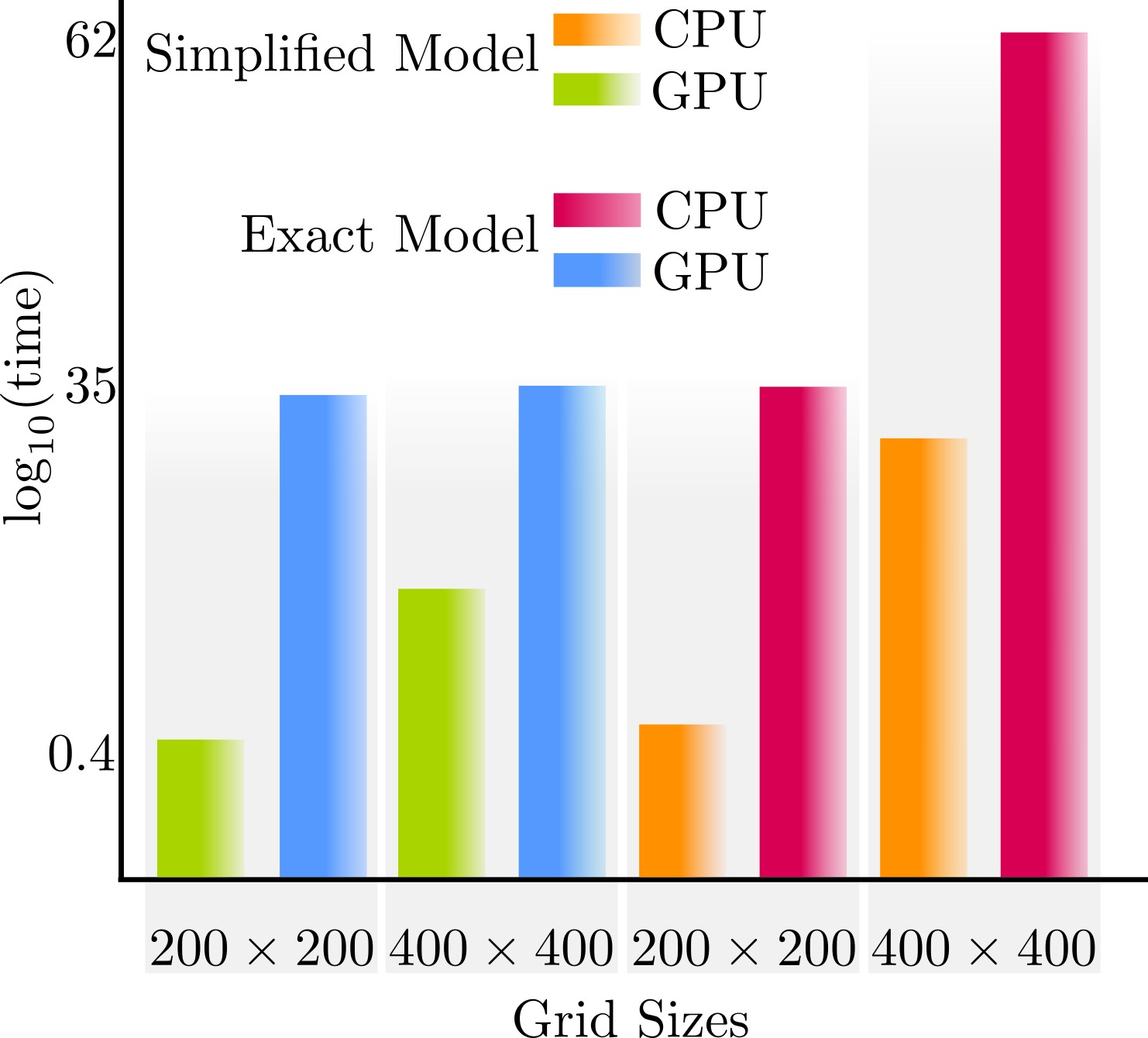}
\caption{Comparison of computational cost between the simplified and exact models for transverse 2D grid sizes of $200 \times 200$ and $400 \times 400$. Simulations were performed in MATLAB R2024a on a system with an Intel(R) Core i7-14700K CPU (3.40 GHz), 128 GB RAM, and an NVIDIA GeForce RTX 4070 GPU, running Windows 11 Enterprise. The runtime corresponds to beam propagation over $200~\mu\mathrm{m}$. The vertical axis shows $\log_{10}$ of the runtime in seconds, illustrating the order-of-magnitude computational speedup achieved by the simplified model.}
\label{fig:time}
\end{figure}
As the input power exceeds the nonlinear threshold, optically induced molecular reorientation progressively reduces the inter-waveguide coupling and partially suppresses diffraction along $x$, yielding a more localized intensity profile. At sufficiently high powers, the nonlinear response compensates for both diffraction mechanisms, resulting in robust self-localization: the beam becomes confined to a single waveguide along $x$ (forming a discrete nematicon) and propagates quasi-nondiffractively along $z$ throughout the entire propagation length. Such behavior is observed in both the exact and simplified models at comparable power levels, as shown in Fig.~\ref{light-dynamics}(c) and (d), respectively.

For larger pitch values ($\Lambda = 16~\mu\mathrm{m}$), the coupling between adjacent waveguides is weak, and thus, even at low powers, the beam remains largely confined to the central waveguide with only modest spreading along $x$. In this regime, the light is effectively localized along $x$ while continuing to diffract along $y$ (Fig.~\ref{light-dynamics}(e) and (f)). As the input power increases beyond the nonlinear threshold, self-localization is achieved in both transverse directions, and the beam propagates quasi-nondiffractively along $z$ in both the exact and simplified models, as shown in Fig.~\ref{light-dynamics}(g) and (h). The only noticeable difference between the two models is a slight shift in the threshold power for self-localization; however, the overall power range for the transition remains consistent.

We note that the input powers required to reach comparable self-localized states differ slightly between the exact and reduced-order models. This systematic shift in the self-localization threshold arises from the neglect of full spatial nonlocality in the reduced-order description, which captures the saturable response but does not include long-range elastic interactions. As a result, the reduced-order model typically requires somewhat higher input powers to achieve the same degree of confinement. Importantly, despite this quantitative shift, the reduced-order model preserves the correct qualitative transitions and overall power scaling of the nonlinear dynamics.

Although the proposed simplified model accounts only for the saturable nonlinear response, it successfully reproduces the same qualitative features of the light dynamics as the exact model.

\subsubsection{Computational Time Benefits}

Fig.~\ref{fig:time} summarizes the computational advantages of the proposed reduced-order model.  
Compared with full numerical relaxation-based simulations, the simplified formulation achieves a reduction in runtime of approximately two orders of magnitude, while maintaining good agreement, within the model’s validity, in the predicted soliton widths, beam dynamics, and localization thresholds.  
This dramatic speedup allows extensive parameter exploration—including variations in pitch, birefringence, beam waist, and input power.

The model's efficiency makes it an effective tool for rapid design and optimization of nonlinear liquid-crystal photonic systems.  
Beyond the computational benefits, the approach preserves the key physical mechanisms of saturable self-focusing and the transition from discrete diffraction to soliton formation, offering both numerical practicality and clear physical insight into light–matter coupling in chiral nematic liquid crystals. The model framework may also inform emerging reconfigurable soft-photonic systems such as LC-integrated metasurfaces~\cite{kang2024liquid} and nonlinear adaptive optics platforms~\cite{Piccardi2013IJMS}.

\section{Conclusions}
In conclusion, we have developed a reduced-order, semi-analytical description of reorientational nonlinearity in chiral nematic liquid crystals that captures the essential features of beam self-focusing and soliton formation while significantly reducing computational cost. The central physical insight of this work is that the chiral pitch acts as the dominant length scale governing effective nonlinear confinement, setting a crossover between delocalized propagation and single-channel self-localization. This behavior is encoded through an effective chiral confinement length $\omega_{\mathrm{ch}}$, which provides a transparent interpretation of the nonlinear response in terms of a pitch-controlled geometric transition. While the reduced-order model neglects full spatial nonlocality and therefore exhibits systematic shifts in the self-localization threshold, it preserves the correct qualitative transitions and overall power scaling observed in full numerical simulations. As such, the model offers a physically interpretable and computationally efficient framework for exploring nonlinear light dynamics in pitch-modulated nematic media within its range of validity. Related experimental observations of such reorientation-driven transitions in chiral nematic systems have been reported previously~\cite{laudyn2014power, laudyn2016quasi, piccardi2024nematonics}.

\appendix
\section{Derivation of the analytical expression for the refractive index}
We start from the approximate expression for Eq.~\eqref{nnn} under the small-anisotropy assumption:
\begin{equation}
    n \approx n_o + (n_e - n_o)\cos^2\theta 
    = n_o + \frac{1}{2}(n_e - n_o)(1 + \cos 2\theta).
\end{equation}
Next, substituting $\theta = \theta_0 - \varphi$ and using the expression for $\varphi$ given by
\begin{equation}
    \varphi = \frac{1}{2}\tan^{-1}\!\left(
    \frac{ I \sin 2\theta_0 }{ \frac{\pi}{2} + I \cos 2\theta_0 }
    \right), 
    \label{phi_equation}
\end{equation}
we find
\begin{align}
    \cos(2\theta) 
    &= \cos(2\theta_0 - 2\varphi) \nonumber \\
    &= \cos 2\theta_0 \cos 2\varphi 
       + \sin 2\theta_0 \sin 2\varphi.
\end{align}

Using the trigonometric identities
\[
\cos(\tan^{-1}x) = \frac{1}{\sqrt{1 + x^2}},
\quad 
\sin(\tan^{-1}x) = \frac{x}{\sqrt{1 + x^2}},
\]
and defining
\[
x = \frac{I \sin 2\theta_0}{\frac{\pi}{2} + I \cos 2\theta_0},
\]
we can simplify $\cos(2\theta)$ as
\begin{equation}
    \cos 2\theta 
    = -\frac{\frac{\pi}{2} \cos 2\theta_0 + I}
    {\sqrt{\left(\frac{\pi}{2}\right)^2 
    + \pi I \cos 2\theta_0 + I^2}}.
\end{equation}

Substituting this result back, the approximate refractive index becomes
\begin{equation}
    n \approx n_o 
    + \frac{1}{2}(n_e - n_o)
    \left[
    1 - 
    \frac{
    \frac{\pi}{2} \cos 2\theta_0 + I
    }{
    \sqrt{
    \left(\frac{\pi}{2}\right)^2 
    + \pi I \cos 2\theta_0 + I^2
    }
    }
    \right].
\end{equation}



\begin{thebibliography}{49}%
\makeatletter
\providecommand \@ifxundefined [1]{%
 \@ifx{#1\undefined}
}%
\providecommand \@ifnum [1]{%
 \ifnum #1\expandafter \@firstoftwo
 \else \expandafter \@secondoftwo
 \fi
}%
\providecommand \@ifx [1]{%
 \ifx #1\expandafter \@firstoftwo
 \else \expandafter \@secondoftwo
 \fi
}%
\providecommand \natexlab [1]{#1}%
\providecommand \enquote  [1]{``#1''}%
\providecommand \bibnamefont  [1]{#1}%
\providecommand \bibfnamefont [1]{#1}%
\providecommand \citenamefont [1]{#1}%
\providecommand \href@noop [0]{\@secondoftwo}%
\providecommand \href [0]{\begingroup \@sanitize@url \@href}%
\providecommand \@href[1]{\@@startlink{#1}\@@href}%
\providecommand \@@href[1]{\endgroup#1\@@endlink}%
\providecommand \@sanitize@url [0]{\catcode `\\12\catcode `\$12\catcode `\&12\catcode `\#12\catcode `\^12\catcode `\_12\catcode `\%12\relax}%
\providecommand \@@startlink[1]{}%
\providecommand \@@endlink[0]{}%
\providecommand \url  [0]{\begingroup\@sanitize@url \@url }%
\providecommand \@url [1]{\endgroup\@href {#1}{\urlprefix }}%
\providecommand \urlprefix  [0]{URL }%
\providecommand \Eprint [0]{\href }%
\providecommand \doibase [0]{https://doi.org/}%
\providecommand \selectlanguage [0]{\@gobble}%
\providecommand \bibinfo  [0]{\@secondoftwo}%
\providecommand \bibfield  [0]{\@secondoftwo}%
\providecommand \translation [1]{[#1]}%
\providecommand \BibitemOpen [0]{}%
\providecommand \bibitemStop [0]{}%
\providecommand \bibitemNoStop [0]{.\EOS\space}%
\providecommand \EOS [0]{\spacefactor3000\relax}%
\providecommand \BibitemShut  [1]{\csname bibitem#1\endcsname}%
\let\auto@bib@innerbib\@empty
\bibitem [{\citenamefont {Akhmanov}\ \emph {et~al.}(1968)\citenamefont {Akhmanov}, \citenamefont {Sukhorukov},\ and\ \citenamefont {Khokhlov}}]{akhmanov1968self}%
  \BibitemOpen
  \bibfield  {author} {\bibinfo {author} {\bibfnamefont {S.~A.}\ \bibnamefont {Akhmanov}}, \bibinfo {author} {\bibfnamefont {A.~P.}\ \bibnamefont {Sukhorukov}},\ and\ \bibinfo {author} {\bibfnamefont {R.~V.}\ \bibnamefont {Khokhlov}},\ }\bibfield  {title} {\bibinfo {title} {Self-focusing and diffraction of light in a nonlinear medium},\ }\href {https://doi.org/10.1070/PU1968v010n05ABEH005849} {\bibfield  {journal} {\bibinfo  {journal} {Soviet Physics Uspekhi}\ }\textbf {\bibinfo {volume} {10}},\ \bibinfo {pages} {609} (\bibinfo {year} {1968})}\BibitemShut {NoStop}%
\bibitem [{\citenamefont {Litvak}(1965)}]{Litvak1965}%
  \BibitemOpen
  \bibfield  {author} {\bibinfo {author} {\bibfnamefont {A.~G.}\ \bibnamefont {Litvak}},\ }\bibfield  {title} {\bibinfo {title} {Self-focusing of electromagnetic waves in a plasma in an intense magnetic field},\ }\href {https://doi.org/10.1007/BF01038281} {\bibfield  {journal} {\bibinfo  {journal} {Soviet Radiophysics}\ }\textbf {\bibinfo {volume} {8}},\ \bibinfo {pages} {828} (\bibinfo {year} {1965})}\BibitemShut {NoStop}%
\bibitem [{\citenamefont {Shukla}\ \emph {et~al.}(1992)\citenamefont {Shukla}, \citenamefont {Stenflo},\ and\ \citenamefont {Borisov}}]{Shukla1992}%
  \BibitemOpen
  \bibfield  {author} {\bibinfo {author} {\bibfnamefont {P.~K.}\ \bibnamefont {Shukla}}, \bibinfo {author} {\bibfnamefont {L.}~\bibnamefont {Stenflo}},\ and\ \bibinfo {author} {\bibfnamefont {N.~D.}\ \bibnamefont {Borisov}},\ }\bibfield  {title} {\bibinfo {title} {Nonlinear interaction of powerful radio waves with the plasma in the earth's lower ionosphere},\ }\href {https://doi.org/10.1029/92JA00728} {\bibfield  {journal} {\bibinfo  {journal} {Journal of Geophysical Research: Space Physics}\ }\textbf {\bibinfo {volume} {97}},\ \bibinfo {pages} {12279} (\bibinfo {year} {1992})}\BibitemShut {NoStop}%
\bibitem [{\citenamefont {Dalfovo}\ \emph {et~al.}(1999)\citenamefont {Dalfovo}, \citenamefont {Giorgini}, \citenamefont {Pitaevskii},\ and\ \citenamefont {Stringari}}]{Dalfovo1999}%
  \BibitemOpen
  \bibfield  {author} {\bibinfo {author} {\bibfnamefont {F.}~\bibnamefont {Dalfovo}}, \bibinfo {author} {\bibfnamefont {S.}~\bibnamefont {Giorgini}}, \bibinfo {author} {\bibfnamefont {L.~P.}\ \bibnamefont {Pitaevskii}},\ and\ \bibinfo {author} {\bibfnamefont {S.}~\bibnamefont {Stringari}},\ }\bibfield  {title} {\bibinfo {title} {Theory of bose–einstein condensation in trapped gases},\ }\href {https://doi.org/10.1103/RevModPhys.71.463} {\bibfield  {journal} {\bibinfo  {journal} {Reviews of Modern Physics}\ }\textbf {\bibinfo {volume} {71}},\ \bibinfo {pages} {463} (\bibinfo {year} {1999})}\BibitemShut {NoStop}%
\bibitem [{\citenamefont {Pitaevskii}\ and\ \citenamefont {Stringari}(2003)}]{Pitaevskii2003}%
  \BibitemOpen
  \bibfield  {author} {\bibinfo {author} {\bibfnamefont {L.~P.}\ \bibnamefont {Pitaevskii}}\ and\ \bibinfo {author} {\bibfnamefont {S.}~\bibnamefont {Stringari}},\ }\href@noop {} {\emph {\bibinfo {title} {Bose–Einstein Condensation}}}\ (\bibinfo  {publisher} {Oxford University Press},\ \bibinfo {year} {2003})\BibitemShut {NoStop}%
\bibitem [{\citenamefont {Achilleos}\ \emph {et~al.}(2015)\citenamefont {Achilleos}, \citenamefont {Richoux}, \citenamefont {Theocharis},\ and\ \citenamefont {Frantzeskakis}}]{Achilleos2015}%
\BibitemOpen
\bibfield {author} {\bibinfo {author} {\bibfnamefont {V.}~\bibnamefont {Achilleos}}, \bibinfo {author} {\bibfnamefont {O.}~\bibnamefont {Richoux}}, \bibinfo {author} {\bibfnamefont {G.}~\bibnamefont {Theocharis}},\ and\ \bibinfo {author} {\bibfnamefont {D.~J.}\ \bibnamefont {Frantzeskakis}},\ }
\bibfield {title} {\bibinfo {title} {Acoustic solitons in waveguides with Helmholtz resonators: Transmission line approach},\ }
\href {https://doi.org/10.1103/PhysRevE.91.023204} {\bibfield {journal} {\bibinfo {journal} {Physical Review E}\ }\textbf {\bibinfo {volume} {91}},\ \bibinfo {pages} {023204} (\bibinfo {year} {2015})}\BibitemShut {NoStop}%
\bibitem [{\citenamefont {Kuznetsov}\ \emph {et~al.}(1986)\citenamefont {Kuznetsov}, \citenamefont {Rubenchik},\ and\ \citenamefont {Zakharov}}]{kuznetsov1986}%
  \BibitemOpen
  \bibfield  {author} {\bibinfo {author} {\bibfnamefont {E.~A.}\ \bibnamefont {Kuznetsov}}, \bibinfo {author} {\bibfnamefont {A.~M.}\ \bibnamefont {Rubenchik}},\ and\ \bibinfo {author} {\bibfnamefont {V.~E.}\ \bibnamefont {Zakharov}},\ }\bibfield  {title} {\bibinfo {title} {Soliton stability in plasmas and hydrodynamics},\ }\href {https://doi.org/10.1016/0370-1573(86)90016-5} {\bibfield  {journal} {\bibinfo  {journal} {Physics Reports}\ }\textbf {\bibinfo {volume} {142}},\ \bibinfo {pages} {103} (\bibinfo {year} {1986})}\BibitemShut {NoStop}%
\bibitem [{\citenamefont {Shukla}\ and\ \citenamefont {Stenflo}(1984)}]{Shukla1984}%
  \BibitemOpen
  \bibfield  {author} {\bibinfo {author} {\bibfnamefont {P.~K.}\ \bibnamefont {Shukla}}\ and\ \bibinfo {author} {\bibfnamefont {L.}~\bibnamefont {Stenflo}},\ }\bibfield  {title} {\bibinfo {title} {Nonlinear propagation of electromagnetic waves in magnetized plasmas},\ }\href {https://doi.org/10.1103/PhysRevA.30.2110} {\bibfield  {journal} {\bibinfo  {journal} {Physical Review A}\ }\textbf {\bibinfo {volume} {30}},\ \bibinfo {pages} {2110} (\bibinfo {year} {1984})}\BibitemShut {NoStop}%
\bibitem [{\citenamefont {Kivshar}\ and\ \citenamefont {Agrawal}(2003)}]{Kivshar2003}%
  \BibitemOpen
  \bibfield  {author} {\bibinfo {author} {\bibfnamefont {Y.~S.}\ \bibnamefont {Kivshar}}\ and\ \bibinfo {author} {\bibfnamefont {G.~P.}\ \bibnamefont {Agrawal}},\ }\href@noop {} {\emph {\bibinfo {title} {Optical Solitons: From Fibers to Photonic Crystals}}}\ (\bibinfo  {publisher} {Academic Press},\ \bibinfo {address} {San Diego},\ \bibinfo {year} {2003})\BibitemShut {NoStop}%
\bibitem [{\citenamefont {Abdullaev}\ \emph {et~al.}(2004)\citenamefont {Abdullaev}, \citenamefont {Gammal},\ and\ \citenamefont {Tomio}}]{Abdullaev2004}%
  \BibitemOpen
  \bibfield  {author} {\bibinfo {author} {\bibfnamefont {F.~K.}\ \bibnamefont {Abdullaev}}, \bibinfo {author} {\bibfnamefont {A.}~\bibnamefont {Gammal}},\ and\ \bibinfo {author} {\bibfnamefont {L.}~\bibnamefont {Tomio}},\ }\bibfield  {title} {\bibinfo {title} {Dynamics of bright matter-wave solitons in a bose–einstein condensate with inhomogeneous scattering length},\ }\href {https://doi.org/10.1088/0953-4075/37/3/009} {\bibfield  {journal} {\bibinfo  {journal} {Journal of Physics B}\ }\textbf {\bibinfo {volume} {37}},\ \bibinfo {pages} {635} (\bibinfo {year} {2004})}\BibitemShut {NoStop}%
\bibitem [{\citenamefont {Wabnitz}(2017)}]{Wabnitz2017}%
  \BibitemOpen
  \bibfield  {author} {\bibinfo {author} {\bibfnamefont {S.}~\bibnamefont {Wabnitz}},\ }\href {https://doi.org/10.1088/978-0-7503-1460-2} {\emph {\bibinfo {title} {Nonlinear guided wave optics}}}\ (\bibinfo  {publisher} {IOP Publishing},\ \bibinfo {year} {2017})\BibitemShut {NoStop}%
\bibitem [{\citenamefont {Sukhorukov}\ and\ \citenamefont {Kivshar}(2002)}]{sukhorukov2002nonlinear}%
  \BibitemOpen
  \bibfield  {author} {\bibinfo {author} {\bibfnamefont {A.~A.}\ \bibnamefont {Sukhorukov}}\ and\ \bibinfo {author} {\bibfnamefont {Y.~S.}\ \bibnamefont {Kivshar}},\ }\bibfield  {title} {\bibinfo {title} {Nonlinear guided waves and spatial solitons in a periodic layered medium},\ }\href {https://doi.org/10.1364/JOSAB.19.000772} {\bibfield  {journal} {\bibinfo  {journal} {Journal of the Optical Society of America B}\ }\textbf {\bibinfo {volume} {19}},\ \bibinfo {pages} {772} (\bibinfo {year} {2002})}\BibitemShut {NoStop}%
\bibitem [{\citenamefont {Bang}\ \emph {et~al.}(1999)\citenamefont {Bang}, \citenamefont {Edmundson},\ and\ \citenamefont {Krolikowski}}]{bang1999collapse}%
  \BibitemOpen
  \bibfield  {author} {\bibinfo {author} {\bibfnamefont {O.}~\bibnamefont {Bang}}, \bibinfo {author} {\bibfnamefont {D.}~\bibnamefont {Edmundson}},\ and\ \bibinfo {author} {\bibfnamefont {W.}~\bibnamefont {Krolikowski}},\ }\bibfield  {title} {\bibinfo {title} {Collapse of incoherent light beams in inertial bulk kerr media},\ }\href {https://doi.org/10.1103/PhysRevLett.83.5479} {\bibfield  {journal} {\bibinfo  {journal} {Physical Review Letters}\ }\textbf {\bibinfo {volume} {83}},\ \bibinfo {pages} {5479} (\bibinfo {year} {1999})}\BibitemShut {NoStop}%
\bibitem [{\citenamefont {Bang}\ \emph {et~al.}(2002)\citenamefont {Bang}, \citenamefont {Krolikowski}, \citenamefont {Wyller},\ and\ \citenamefont {Rasmussen}}]{bang2002collapse}%
  \BibitemOpen
  \bibfield  {author} {\bibinfo {author} {\bibfnamefont {O.}~\bibnamefont {Bang}}, \bibinfo {author} {\bibfnamefont {W.}~\bibnamefont {Krolikowski}}, \bibinfo {author} {\bibfnamefont {J.}~\bibnamefont {Wyller}},\ and\ \bibinfo {author} {\bibfnamefont {J.~J.}\ \bibnamefont {Rasmussen}},\ }\bibfield  {title} {\bibinfo {title} {Collapse arrest and soliton stabilization in nonlocal nonlinear media},\ }\href {https://doi.org/10.1103/PhysRevE.66.046619} {\bibfield  {journal} {\bibinfo  {journal} {Physical Review E}\ }\textbf {\bibinfo {volume} {66}},\ \bibinfo {pages} {046619} (\bibinfo {year} {2002})}\BibitemShut {NoStop}%
\bibitem [{\citenamefont {Snyder}\ and\ \citenamefont {Mitchell}(1997)}]{Snyder1997}%
  \BibitemOpen
  \bibfield  {author} {\bibinfo {author} {\bibfnamefont {A.~W.}\ \bibnamefont {Snyder}}\ and\ \bibinfo {author} {\bibfnamefont {D.~J.}\ \bibnamefont {Mitchell}},\ }\bibfield  {title} {\bibinfo {title} {Accessible solitons},\ }\href {https://doi.org/10.1126/science.276.5318.1538} {\bibfield  {journal} {\bibinfo  {journal} {Science}\ }\textbf {\bibinfo {volume} {276}},\ \bibinfo {pages} {1538} (\bibinfo {year} {1997})}\BibitemShut {NoStop}%
\bibitem [{\citenamefont {Krolikowski}\ \emph {et~al.}(2001)\citenamefont {Krolikowski}, \citenamefont {Bang}, \citenamefont {Rasmussen},\ and\ \citenamefont {Wyller}}]{Krolikowski2001}%
  \BibitemOpen
  \bibfield  {author} {\bibinfo {author} {\bibfnamefont {W.}~\bibnamefont {Krolikowski}}, \bibinfo {author} {\bibfnamefont {O.}~\bibnamefont {Bang}}, \bibinfo {author} {\bibfnamefont {J.~J.}\ \bibnamefont {Rasmussen}},\ and\ \bibinfo {author} {\bibfnamefont {J.}~\bibnamefont {Wyller}},\ }\bibfield  {title} {\bibinfo {title} {Modulational instability in nonlocal nonlinear kerr media},\ }\href {https://doi.org/10.1103/PhysRevE.64.016612} {\bibfield  {journal} {\bibinfo  {journal} {Physical Review E}\ }\textbf {\bibinfo {volume} {64}},\ \bibinfo {pages} {016612} (\bibinfo {year} {2001})}\BibitemShut {NoStop}%
\bibitem [{\citenamefont {Conti}\ \emph {et~al.}(2003)\citenamefont {Conti}, \citenamefont {Peccianti},\ and\ \citenamefont {Assanto}}]{conti2003route}%
  \BibitemOpen
  \bibfield  {author} {\bibinfo {author} {\bibfnamefont {C.}~\bibnamefont {Conti}}, \bibinfo {author} {\bibfnamefont {M.}~\bibnamefont {Peccianti}},\ and\ \bibinfo {author} {\bibfnamefont {G.}~\bibnamefont {Assanto}},\ }\bibfield  {title} {\bibinfo {title} {Route to nonlocality and observation of accessible solitons},\ }\href {https://doi.org/10.1103/PhysRevLett.91.073901} {\bibfield  {journal} {\bibinfo  {journal} {Physical Review Letters}\ }\textbf {\bibinfo {volume} {91}},\ \bibinfo {pages} {073901} (\bibinfo {year} {2003})}\BibitemShut {NoStop}%
\bibitem [{\citenamefont {Peccianti}\ \emph {et~al.}(2006)\citenamefont {Peccianti}, \citenamefont {Dyadyusha}, \citenamefont {Kaczmarek},\ and\ \citenamefont {Assanto}}]{peccianti2006nonlinear}%
  \BibitemOpen
  \bibfield  {author} {\bibinfo {author} {\bibfnamefont {M.}~\bibnamefont {Peccianti}}, \bibinfo {author} {\bibfnamefont {A.}~\bibnamefont {Dyadyusha}}, \bibinfo {author} {\bibfnamefont {M.}~\bibnamefont {Kaczmarek}},\ and\ \bibinfo {author} {\bibfnamefont {G.}~\bibnamefont {Assanto}},\ }\bibfield  {title} {\bibinfo {title} {Tunable refraction and reflection of self-confined light beams},\ }\href {https://doi.org/10.1038/nphys427} {\bibfield  {journal} {\bibinfo  {journal} {Nature Physics}\ }\textbf {\bibinfo {volume} {2}},\ \bibinfo {pages} {737} (\bibinfo {year} {2006})}\BibitemShut {NoStop}%
\bibitem [{\citenamefont {Assanto}\ and\ \citenamefont {Karpierz}(2009)}]{assanto2009nematicons}%
\BibitemOpen
\bibfield {author} {\bibinfo {author} {\bibfnamefont {G.}~\bibnamefont {Assanto}}\ and\ \bibinfo {author} {\bibfnamefont {M.~A.}\ \bibnamefont {Karpierz}},\ }
\bibfield {title} {\bibinfo {title} {Nematicons: self-localised beams in nematic liquid crystals},\ }
\href {https://doi.org/10.1080/02678290903033441} {\bibfield {journal} {\bibinfo {journal} {Liquid Crystals}\ }\textbf {\bibinfo {volume} {36}},\ \bibinfo {pages} {1161--1172} (\bibinfo {year} {2009})}\BibitemShut {NoStop}%
\bibitem [{\citenamefont {Peccianti}\ and\ \citenamefont {Assanto}(2012)}]{peccianti2012nematicons}%
  \BibitemOpen
  \bibfield  {author} {\bibinfo {author} {\bibfnamefont {M.}~\bibnamefont {Peccianti}}\ and\ \bibinfo {author} {\bibfnamefont {G.}~\bibnamefont {Assanto}},\ }\bibfield  {title} {\bibinfo {title} {Nematicons},\ }\href {https://doi.org/10.1016/j.physrep.2012.02.004} {\bibfield  {journal} {\bibinfo  {journal} {Physics Reports}\ }\textbf {\bibinfo {volume} {516}},\ \bibinfo {pages} {147} (\bibinfo {year} {2012})}\BibitemShut {NoStop}%
\bibitem [{\citenamefont {Braun}\ \emph {et~al.}(1993)\citenamefont {Braun}, \citenamefont {Faucheux},\ and\ \citenamefont {Libchaber}}]{braun1993}%
  \BibitemOpen
  \bibfield  {author} {\bibinfo {author} {\bibfnamefont {E.}~\bibnamefont {Braun}}, \bibinfo {author} {\bibfnamefont {L.~P.}\ \bibnamefont {Faucheux}},\ and\ \bibinfo {author} {\bibfnamefont {A.}~\bibnamefont {Libchaber}},\ }\bibfield  {title} {\bibinfo {title} {Strong self-focusing in nematic liquid crystals},\ }\href {https://doi.org/10.1103/PhysRevA.48.611} {\bibfield  {journal} {\bibinfo  {journal} {Physical Review A}\ }\textbf {\bibinfo {volume} {48}},\ \bibinfo {pages} {611} (\bibinfo {year} {1993})}\BibitemShut {NoStop}%
\bibitem [{\citenamefont {Conti}\ \emph {et~al.}(2004)\citenamefont {Conti}, \citenamefont {Peccianti},\ and\ \citenamefont {Assanto}}]{conti2004observation}%
  \BibitemOpen
  \bibfield  {author} {\bibinfo {author} {\bibfnamefont {C.}~\bibnamefont {Conti}}, \bibinfo {author} {\bibfnamefont {M.}~\bibnamefont {Peccianti}},\ and\ \bibinfo {author} {\bibfnamefont {G.}~\bibnamefont {Assanto}},\ }\bibfield  {title} {\bibinfo {title} {Observation of optical spatial solitons in a highly nonlocal medium},\ }\href {https://doi.org/10.1103/PhysRevLett.92.113902} {\bibfield  {journal} {\bibinfo  {journal} {Physical Review Letters}\ }\textbf {\bibinfo {volume} {92}},\ \bibinfo {pages} {113902} (\bibinfo {year} {2004})}\BibitemShut {NoStop}%
\bibitem [{\citenamefont {Piccardi}\ and\ \citenamefont {Assanto}(2024)}]{piccardi2024nematonics}%
  \BibitemOpen
  \bibfield  {author} {\bibinfo {author} {\bibfnamefont {A.}~\bibnamefont {Piccardi}}\ and\ \bibinfo {author} {\bibfnamefont {G.}~\bibnamefont {Assanto}},\ }\bibfield  {title} {\bibinfo {title} {Nematonics: from physics to photonics of reorientational solitons},\ }\href {https://doi.org/10.1080/02678292.2024.2314628} {\bibfield  {journal} {\bibinfo  {journal} {Liquid Crystals}\ }\textbf {\bibinfo {volume} {51}},\ \bibinfo {pages} {2252} (\bibinfo {year} {2024})}\BibitemShut {NoStop}%
\bibitem [{\citenamefont {Hutsebaut}\ \emph {et~al.}(2004)\citenamefont {Hutsebaut}, \citenamefont {Cambournac}, \citenamefont {Haelterman}, \citenamefont {Adamski},\ and\ \citenamefont {Neyts}}]{hutsebaut2004single}%
  \BibitemOpen
  \bibfield  {author} {\bibinfo {author} {\bibfnamefont {X.}~\bibnamefont {Hutsebaut}}, \bibinfo {author} {\bibfnamefont {C.}~\bibnamefont {Cambournac}}, \bibinfo {author} {\bibfnamefont {M.}~\bibnamefont {Haelterman}}, \bibinfo {author} {\bibfnamefont {A.}~\bibnamefont {Adamski}},\ and\ \bibinfo {author} {\bibfnamefont {K.}~\bibnamefont {Neyts}},\ }\bibfield  {title} {\bibinfo {title} {Single-component higher-order mode solitons in liquid crystals},\ }\href {https://doi.org/10.1016/j.optcom.2004.01.040} {\bibfield  {journal} {\bibinfo  {journal} {Optics Communications}\ }\textbf {\bibinfo {volume} {233}},\ \bibinfo {pages} {211} (\bibinfo {year} {2004})}\BibitemShut {NoStop}%
\bibitem [{\citenamefont {Laudyn}\ \emph {et~al.}(2015)\citenamefont {Laudyn}, \citenamefont {Kwasny}, \citenamefont {Piccardi}, \citenamefont {Karpierz}, \citenamefont {Dabrowski}, \citenamefont {Chojnowska}, \citenamefont {Alberucci},\ and\ \citenamefont {Assanto}}]{laudyn2015nonlinear}%
  \BibitemOpen
  \bibfield  {author} {\bibinfo {author} {\bibfnamefont {U.~A.}\ \bibnamefont {Laudyn}}, \bibinfo {author} {\bibfnamefont {M.}~\bibnamefont {Kwasny}}, \bibinfo {author} {\bibfnamefont {A.}~\bibnamefont {Piccardi}}, \bibinfo {author} {\bibfnamefont {M.~A.}\ \bibnamefont {Karpierz}}, \bibinfo {author} {\bibfnamefont {R.}~\bibnamefont {Dabrowski}}, \bibinfo {author} {\bibfnamefont {O.}~\bibnamefont {Chojnowska}}, \bibinfo {author} {\bibfnamefont {A.}~\bibnamefont {Alberucci}},\ and\ \bibinfo {author} {\bibfnamefont {G.}~\bibnamefont {Assanto}},\ }\bibfield  {title} {\bibinfo {title} {Nonlinear competition in nematicon propagation},\ }\href {https://doi.org/10.1364/OL.40.005235} {\bibfield  {journal} {\bibinfo  {journal} {Optics Letters}\ }\textbf {\bibinfo {volume} {40}},\ \bibinfo {pages} {5235} (\bibinfo {year} {2015})}\BibitemShut {NoStop}%
\bibitem [{\citenamefont {Warenghem}\ \emph {et~al.}(2008)\citenamefont {Warenghem}, \citenamefont {Blach},\ and\ \citenamefont {Henninot}}]{warenghem2008thermo}%
  \BibitemOpen
  \bibfield  {author} {\bibinfo {author} {\bibfnamefont {M.}~\bibnamefont {Warenghem}}, \bibinfo {author} {\bibfnamefont {J.~F.}\ \bibnamefont {Blach}},\ and\ \bibinfo {author} {\bibfnamefont {J.~F.}\ \bibnamefont {Henninot}},\ }\bibfield  {title} {\bibinfo {title} {Thermo-nematicon: an unnatural coexistence of solitons in liquid crystals?},\ }\href {https://doi.org/10.1364/JOSAB.25.001882} {\bibfield  {journal} {\bibinfo  {journal} {Journal of the Optical Society of America B}\ }\textbf {\bibinfo {volume} {25}},\ \bibinfo {pages} {1882} (\bibinfo {year} {2008})}\BibitemShut {NoStop}%
\bibitem [{\citenamefont {Jung}\ \emph {et~al.}(2017)\citenamefont {Jung}, \citenamefont {Krolikowski}, \citenamefont {Laudyn}, \citenamefont {Trippenbach},\ and\ \citenamefont {Karpierz}}]{jung2017supermode}%
  \BibitemOpen
  \bibfield  {author} {\bibinfo {author} {\bibfnamefont {P.~S.}\ \bibnamefont {Jung}}, \bibinfo {author} {\bibfnamefont {W.}~\bibnamefont {Krolikowski}}, \bibinfo {author} {\bibfnamefont {U.~A.}\ \bibnamefont {Laudyn}}, \bibinfo {author} {\bibfnamefont {M.}~\bibnamefont {Trippenbach}},\ and\ \bibinfo {author} {\bibfnamefont {M.~A.}\ \bibnamefont {Karpierz}},\ }\bibfield  {title} {\bibinfo {title} {Supermode spatial optical solitons in liquid crystals with competing nonlinearities},\ }\href {https://doi.org/10.1103/PhysRevA.95.023820} {\bibfield  {journal} {\bibinfo  {journal} {Physical Review A}\ }\textbf {\bibinfo {volume} {95}},\ \bibinfo {pages} {023820} (\bibinfo {year} {2017})}\BibitemShut {NoStop}%
\bibitem [{\citenamefont {Batarseh}\ \emph {et~al.}(2025)\citenamefont {Batarseh}, \citenamefont {Zakeri}, \citenamefont {Blanco-Redondo}, \citenamefont {Trippenbach}, \citenamefont {Hagan}, \citenamefont {Krolikowski},\ and\ \citenamefont {Jung}}]{batarseh2025crossover}%
  \BibitemOpen
  \bibfield  {author} {\bibinfo {author} {\bibfnamefont {A.~B.}\ \bibnamefont {Batarseh}}, \bibinfo {author} {\bibfnamefont {M.~J.}\ \bibnamefont {Zakeri}}, \bibinfo {author} {\bibfnamefont {A.}~\bibnamefont {Blanco-Redondo}}, \bibinfo {author} {\bibfnamefont {M.}~\bibnamefont {Trippenbach}}, \bibinfo {author} {\bibfnamefont {D.}~\bibnamefont {Hagan}}, \bibinfo {author} {\bibfnamefont {W.}~\bibnamefont {Krolikowski}},\ and\ \bibinfo {author} {\bibfnamefont {P.~S.}\ \bibnamefont {Jung}},\ }\bibfield  {title} {\bibinfo {title} {Crossover from single to two-peak fundamental solitons in nonlocal nonlinear media},\ }\href {https://doi.org/10.1016/j.wavemoti.2024.103445} {\bibfield  {journal} {\bibinfo  {journal} {Wave Motion}\ }\textbf {\bibinfo {volume} {133}},\ \bibinfo {pages} {103445} (\bibinfo {year} {2025})}\BibitemShut {NoStop}%
\bibitem [{\citenamefont {Jung}\ \emph {et~al.}(2024)\citenamefont {Jung}, \citenamefont {Zakeri}, \citenamefont {Ramaniuk}, \citenamefont {Blanco-Redondo}, \citenamefont {Hagan}, \citenamefont {Dogariu}, \citenamefont {Christodoulides}, \citenamefont {Assanto}, \citenamefont {Krolikowski},\ and\ \citenamefont {Trippenbach}}]{jung2024giant}%
  \BibitemOpen
  \bibfield  {author} {\bibinfo {author} {\bibfnamefont {P.~S.}\ \bibnamefont {Jung}}, \bibinfo {author} {\bibfnamefont {M.~J.}\ \bibnamefont {Zakeri}}, \bibinfo {author} {\bibfnamefont {A.}~\bibnamefont {Ramaniuk}}, \bibinfo {author} {\bibfnamefont {A.}~\bibnamefont {Blanco-Redondo}}, \bibinfo {author} {\bibfnamefont {D.~J.}\ \bibnamefont {Hagan}}, \bibinfo {author} {\bibfnamefont {A.}~\bibnamefont {Dogariu}}, \bibinfo {author} {\bibfnamefont {D.~N.}\ \bibnamefont {Christodoulides}}, \bibinfo {author} {\bibfnamefont {G.}~\bibnamefont {Assanto}}, \bibinfo {author} {\bibfnamefont {W.}~\bibnamefont {Krolikowski}},\ and\ \bibinfo {author} {\bibfnamefont {M.}~\bibnamefont {Trippenbach}},\ }\bibfield  {title} {\bibinfo {title} {Giant oscillations of vector solitons driven by nonreciprocal interactions},\ }in\ \href {https://doi.org/10.1364/CLEO{\_}AT.2024.JW2A.203} {\emph {\bibinfo {booktitle} {CLEO: Conference on Lasers and Electro-Optics}}}\ (\bibinfo  {publisher} {Optica Publishing Group},\ \bibinfo {address}
  {Charlotte, North Carolina},\ \bibinfo {year} {2024})\ p.\ \bibinfo {pages} {JW2A.203}\BibitemShut {NoStop}%
\bibitem [{\citenamefont {Alberucci}\ \emph {et~al.}(2009)\citenamefont {Alberucci}, \citenamefont {Assanto}, \citenamefont {Buccoliero}, \citenamefont {Desyatnikov}, \citenamefont {Marchant},\ and\ \citenamefont {Smyth}}]{Alberucci2009}%
  \BibitemOpen
  \bibfield  {author} {\bibinfo {author} {\bibfnamefont {A.}~\bibnamefont {Alberucci}}, \bibinfo {author} {\bibfnamefont {G.}~\bibnamefont {Assanto}}, \bibinfo {author} {\bibfnamefont {D.}~\bibnamefont {Buccoliero}}, \bibinfo {author} {\bibfnamefont {A.~S.}\ \bibnamefont {Desyatnikov}}, \bibinfo {author} {\bibfnamefont {T.~R.}\ \bibnamefont {Marchant}},\ and\ \bibinfo {author} {\bibfnamefont {N.~F.}\ \bibnamefont {Smyth}},\ }\bibfield  {title} {\bibinfo {title} {Modulation analysis of boundary-induced motion of optical solitary waves in a nematic liquid crystal},\ }\href {https://doi.org/10.1103/PhysRevA.79.043816} {\bibfield  {journal} {\bibinfo  {journal} {Physical Review A}\ }\textbf {\bibinfo {volume} {79}},\ \bibinfo {pages} {043816} (\bibinfo {year} {2009})}\BibitemShut {NoStop}%
\bibitem [{\citenamefont {de~Gennes}\ and\ \citenamefont {Prost}(1993)}]{DeGennes1993}%
  \BibitemOpen
  \bibfield  {author} {\bibinfo {author} {\bibfnamefont {P.~G.}\ \bibnamefont {de~Gennes}}\ and\ \bibinfo {author} {\bibfnamefont {J.}~\bibnamefont {Prost}},\ }\href@noop {} {\emph {\bibinfo {title} {The Physics of Liquid Crystals}}},\ \bibinfo {edition} {2nd}\ ed.\ (\bibinfo  {publisher} {Oxford University Press},\ \bibinfo {year} {1993})\BibitemShut {NoStop}%
\bibitem [{\citenamefont {Oswald}\ \emph {et~al.}(2005)\citenamefont {Oswald}, \citenamefont {Pieranski}, \citenamefont {Goodby},\ and\ \citenamefont {Gray}}]{oswald2005nematic}%
  \BibitemOpen
  \bibfield  {author} {\bibinfo {author} {\bibfnamefont {P.}~\bibnamefont {Oswald}}, \bibinfo {author} {\bibfnamefont {P.}~\bibnamefont {Pieranski}}, \bibinfo {author} {\bibfnamefont {J.~W.}\ \bibnamefont {Goodby}},\ and\ \bibinfo {author} {\bibfnamefont {G.~W.}\ \bibnamefont {Gray}},\ }\href@noop {} {\emph {\bibinfo {title} {Nematic and Cholesteric Liquid Crystals}}}\ (\bibinfo  {publisher} {Taylor \& Francis},\ \bibinfo {address} {Boca Raton},\ \bibinfo {year} {2005})\BibitemShut {NoStop}%
\bibitem [{\citenamefont {Laudyn}\ \emph {et~al.}(2009)\citenamefont {Laudyn}, \citenamefont {Kwasny},\ and\ \citenamefont {Karpierz}}]{laudyn2009nematicons}%
  \BibitemOpen
  \bibfield  {author} {\bibinfo {author} {\bibfnamefont {U.~A.}\ \bibnamefont {Laudyn}}, \bibinfo {author} {\bibfnamefont {M.}~\bibnamefont {Kwasny}},\ and\ \bibinfo {author} {\bibfnamefont {M.~A.}\ \bibnamefont {Karpierz}},\ }\bibfield  {title} {\bibinfo {title} {Nematicons in chiral nematic liquid crystals},\ }\href {https://doi.org/10.1063/1.3093498} {\bibfield  {journal} {\bibinfo  {journal} {Applied Physics Letters}\ }\textbf {\bibinfo {volume} {94}},\ \bibinfo {pages} {091110} (\bibinfo {year} {2009})}\BibitemShut {NoStop}%
\bibitem [{\citenamefont {Laudyn}\ \emph {et~al.}(2014)\citenamefont {Laudyn}, \citenamefont {Jung}, \citenamefont {Zegad{\l}o}, \citenamefont {Karpierz},\ and\ \citenamefont {Assanto}}]{laudyn2014power}%
  \BibitemOpen
  \bibfield  {author} {\bibinfo {author} {\bibfnamefont {U.~A.}\ \bibnamefont {Laudyn}}, \bibinfo {author} {\bibfnamefont {P.}~\bibnamefont {Jung}}, \bibinfo {author} {\bibfnamefont {K.~B.}\ \bibnamefont {Zegad{\l}o}}, \bibinfo {author} {\bibfnamefont {M.~A.}\ \bibnamefont {Karpierz}},\ and\ \bibinfo {author} {\bibfnamefont {G.}~\bibnamefont {Assanto}},\ }\bibfield  {title} {\bibinfo {title} {Power-induced evolution and increased dimensionality of nonlinear modes in reorientational soft matter},\ }\href {https://doi.org/10.1364/OL.39.006399} {\bibfield  {journal} {\bibinfo  {journal} {Optics Letters}\ }\textbf {\bibinfo {volume} {39}},\ \bibinfo {pages} {6399} (\bibinfo {year} {2014})}\BibitemShut {NoStop}%
\bibitem [{\citenamefont {Laudyn}\ \emph {et~al.}(2016)\citenamefont {Laudyn}, \citenamefont {Jung}, \citenamefont {Karpierz},\ and\ \citenamefont {Assanto}}]{laudyn2016quasi}%
  \BibitemOpen
  \bibfield  {author} {\bibinfo {author} {\bibfnamefont {U.~A.}\ \bibnamefont {Laudyn}}, \bibinfo {author} {\bibfnamefont {P.}~\bibnamefont {Jung}}, \bibinfo {author} {\bibfnamefont {M.~A.}\ \bibnamefont {Karpierz}},\ and\ \bibinfo {author} {\bibfnamefont {G.}~\bibnamefont {Assanto}},\ }\bibfield  {title} {\bibinfo {title} {Quasi two-dimensional astigmatic solitons in soft chiral metastructures},\ }\href {https://doi.org/10.1038/srep22923} {\bibfield  {journal} {\bibinfo  {journal} {Scientific Reports}\ }\textbf {\bibinfo {volume} {6}},\ \bibinfo {pages} {22923} (\bibinfo {year} {2016})}\BibitemShut {NoStop}%
\bibitem [{\citenamefont {Ma}\ \emph {et~al.}(2022)\citenamefont {Ma}, \citenamefont {Li}, \citenamefont {Pan}, \citenamefont {Ji}, \citenamefont {Jiang}, \citenamefont {Zheng}, \citenamefont {Wang}, \citenamefont {Wang}, \citenamefont {Li},\ and\ \citenamefont {Lu}}]{ma2022self}%
  \BibitemOpen
  \bibfield  {author} {\bibinfo {author} {\bibfnamefont {L.-L.}\ \bibnamefont {Ma}}, \bibinfo {author} {\bibfnamefont {C.-Y.}\ \bibnamefont {Li}}, \bibinfo {author} {\bibfnamefont {J.-T.}\ \bibnamefont {Pan}}, \bibinfo {author} {\bibfnamefont {Y.-E.}\ \bibnamefont {Ji}}, \bibinfo {author} {\bibfnamefont {C.}~\bibnamefont {Jiang}}, \bibinfo {author} {\bibfnamefont {R.}~\bibnamefont {Zheng}}, \bibinfo {author} {\bibfnamefont {Z.-Y.}\ \bibnamefont {Wang}}, \bibinfo {author} {\bibfnamefont {Y.}~\bibnamefont {Wang}}, \bibinfo {author} {\bibfnamefont {B.-X.}\ \bibnamefont {Li}},\ and\ \bibinfo {author} {\bibfnamefont {Y.-Q.}\ \bibnamefont {Lu}},\ }\bibfield  {title} {\bibinfo {title} {Self-assembled liquid crystal architectures for soft matter photonics},\ }\href {https://doi.org/10.1038/s41377-022-00930-5} {\bibfield  {journal} {\bibinfo  {journal} {Light: Science \& Applications}\ }\textbf {\bibinfo {volume} {11}},\ \bibinfo {pages} {270} (\bibinfo {year} {2022})}\BibitemShut {NoStop}%
\bibitem [{\citenamefont {Kang}\ \emph {et~al.}(2024)\citenamefont {Kang}, \citenamefont {Heo}, \citenamefont {Yang}, \citenamefont {Seong}, \citenamefont {Kim}, \citenamefont {Kim},\ and\ \citenamefont {Rho}}]{kang2024liquid}%
  \BibitemOpen
  \bibfield  {author} {\bibinfo {author} {\bibfnamefont {D.}~\bibnamefont {Kang}}, \bibinfo {author} {\bibfnamefont {H.}~\bibnamefont {Heo}}, \bibinfo {author} {\bibfnamefont {Y.}~\bibnamefont {Yang}}, \bibinfo {author} {\bibfnamefont {J.}~\bibnamefont {Seong}}, \bibinfo {author} {\bibfnamefont {H.}~\bibnamefont {Kim}}, \bibinfo {author} {\bibfnamefont {J.}~\bibnamefont {Kim}},\ and\ \bibinfo {author} {\bibfnamefont {J.}~\bibnamefont {Rho}},\ }\bibfield  {title} {\bibinfo {title} {Liquid crystal-integrated metasurfaces for an active photonic platform},\ }\href {https://doi.org/10.29026/oea.2024.230216} {\bibfield  {journal} {\bibinfo  {journal} {Opto-Electronic Advances}\ }\textbf {\bibinfo {volume} {7}},\ \bibinfo {pages} {230216} (\bibinfo {year} {2024})}\BibitemShut {NoStop}%
\bibitem [{\citenamefont {Zhang}\ \emph {et~al.}(2023)\citenamefont {Zhang}, \citenamefont {Zhang}, \citenamefont {Han}, \citenamefont {Yang}, \citenamefont {Li}, \citenamefont {Song}, \citenamefont {Wang},\ and\ \citenamefont {Zhu}}]{zhang2023advanced}%
  \BibitemOpen
  \bibfield  {author} {\bibinfo {author} {\bibfnamefont {R.}~\bibnamefont {Zhang}}, \bibinfo {author} {\bibfnamefont {Z.}~\bibnamefont {Zhang}}, \bibinfo {author} {\bibfnamefont {J.}~\bibnamefont {Han}}, \bibinfo {author} {\bibfnamefont {L.}~\bibnamefont {Yang}}, \bibinfo {author} {\bibfnamefont {J.}~\bibnamefont {Li}}, \bibinfo {author} {\bibfnamefont {Z.}~\bibnamefont {Song}}, \bibinfo {author} {\bibfnamefont {T.}~\bibnamefont {Wang}},\ and\ \bibinfo {author} {\bibfnamefont {J.}~\bibnamefont {Zhu}},\ }\bibfield  {title} {\bibinfo {title} {Advanced liquid crystal-based switchable optical devices for light protection applications: principles and strategies},\ }\href {https://doi.org/10.1038/s41377-022-01032-y} {\bibfield  {journal} {\bibinfo  {journal} {Light: Science \& Applications}\ }\textbf {\bibinfo {volume} {12}},\ \bibinfo {pages} {11} (\bibinfo {year} {2023})}\BibitemShut {NoStop}%
\bibitem [{\citenamefont {Yang}\ \emph {et~al.}(2021)\citenamefont {Yang}, \citenamefont {Wang}, \citenamefont {Yang},\ and\ \citenamefont {Li}}]{yang20213d}%
  \BibitemOpen
  \bibfield  {author} {\bibinfo {author} {\bibfnamefont {Y.}~\bibnamefont {Yang}}, \bibinfo {author} {\bibfnamefont {L.}~\bibnamefont {Wang}}, \bibinfo {author} {\bibfnamefont {H.}~\bibnamefont {Yang}},\ and\ \bibinfo {author} {\bibfnamefont {Q.}~\bibnamefont {Li}},\ }\bibfield  {title} {\bibinfo {title} {3d chiral photonic nanostructures based on blue-phase liquid crystals},\ }\href {https://doi.org/10.1002/smsc.202100007} {\bibfield  {journal} {\bibinfo  {journal} {Small Science}\ }\textbf {\bibinfo {volume} {1}},\ \bibinfo {pages} {2100007} (\bibinfo {year} {2021})}\BibitemShut {NoStop}%
\bibitem [{\citenamefont {Perumbilavil}\ \emph {et~al.}(2019)\citenamefont {Perumbilavil}, \citenamefont {Kauranen},\ and\ \citenamefont {Assanto}}]{perumbilavil2019spatiospectral}%
  \BibitemOpen
  \bibfield  {author} {\bibinfo {author} {\bibfnamefont {S.}~\bibnamefont {Perumbilavil}}, \bibinfo {author} {\bibfnamefont {M.}~\bibnamefont {Kauranen}},\ and\ \bibinfo {author} {\bibfnamefont {G.}~\bibnamefont {Assanto}},\ }\bibfield  {title} {\bibinfo {title} {Spatiospectral features of a soliton-assisted random laser in liquid crystals},\ }\href {https://doi.org/10.1364/OL.44.003574} {\bibfield  {journal} {\bibinfo  {journal} {Optics Letters}\ }\textbf {\bibinfo {volume} {44}},\ \bibinfo {pages} {3574} (\bibinfo {year} {2019})}\BibitemShut {NoStop}%
\bibitem [{\citenamefont {Assanto}\ and\ \citenamefont {Peccianti}(2003)}]{assanto2003spatial}%
  \BibitemOpen
  \bibfield  {author} {\bibinfo {author} {\bibfnamefont {G.}~\bibnamefont {Assanto}}\ and\ \bibinfo {author} {\bibfnamefont {M.}~\bibnamefont {Peccianti}},\ }\bibfield  {title} {\bibinfo {title} {Spatial solitons in nematic liquid crystals},\ }\href {https://doi.org/10.1109/JQE.2002.806185} {\bibfield  {journal} {\bibinfo  {journal} {IEEE Journal of Quantum Electronics}\ }\textbf {\bibinfo {volume} {39}},\ \bibinfo {pages} {13} (\bibinfo {year} {2003})}\BibitemShut {NoStop}%
\bibitem [{\citenamefont {Christodoulides}\ \emph {et~al.}(2003)\citenamefont {Christodoulides}, \citenamefont {Lederer},\ and\ \citenamefont {Silberberg}}]{christodoulides2003discretizing}%
  \BibitemOpen
  \bibfield  {author} {\bibinfo {author} {\bibfnamefont {D.~N.}\ \bibnamefont {Christodoulides}}, \bibinfo {author} {\bibfnamefont {F.}~\bibnamefont {Lederer}},\ and\ \bibinfo {author} {\bibfnamefont {Y.}~\bibnamefont {Silberberg}},\ }\bibfield  {title} {\bibinfo {title} {Discretizing light behaviour in linear and nonlinear waveguide lattices},\ }\href {https://doi.org/10.1038/nature01936} {\bibfield  {journal} {\bibinfo  {journal} {Nature}\ }\textbf {\bibinfo {volume} {424}},\ \bibinfo {pages} {817} (\bibinfo {year} {2003})}\BibitemShut {NoStop}%
\bibitem [{\citenamefont {Fratalocchi}\ \emph {et~al.}(2004)\citenamefont {Fratalocchi}, \citenamefont {Assanto}, \citenamefont {Brzdąkiewicz},\ and\ \citenamefont {Karpierz}}]{fratalocchi2004discrete}%
  \BibitemOpen
  \bibfield  {author} {\bibinfo {author} {\bibfnamefont {A.}~\bibnamefont {Fratalocchi}}, \bibinfo {author} {\bibfnamefont {G.}~\bibnamefont {Assanto}}, \bibinfo {author} {\bibfnamefont {K.~A.}\ \bibnamefont {Brzdąkiewicz}},\ and\ \bibinfo {author} {\bibfnamefont {M.~A.}\ \bibnamefont {Karpierz}},\ }\bibfield  {title} {\bibinfo {title} {Discrete propagation and spatial solitons in nematic liquid crystals},\ }\href {https://doi.org/10.1364/OL.29.001530} {\bibfield  {journal} {\bibinfo  {journal} {Optics Letters}\ }\textbf {\bibinfo {volume} {29}},\ \bibinfo {pages} {1530} (\bibinfo {year} {2004})}\BibitemShut {NoStop}%
\bibitem [{\citenamefont {Bregar}\ \emph {et~al.}(2018)\citenamefont {Bregar}, \citenamefont {{\v S}timulak},\ and\ \citenamefont {Ravnik}}]{Bregar2018}%
  \BibitemOpen
  \bibfield  {author} {\bibinfo {author} {\bibfnamefont {A.}~\bibnamefont {Bregar}}, \bibinfo {author} {\bibfnamefont {M.}~\bibnamefont {{\v S}timulak}},\ and\ \bibinfo {author} {\bibfnamefont {M.}~\bibnamefont {Ravnik}},\ }\bibfield  {title} {\bibinfo {title} {Photonic properties of heliconical liquid crystals},\ }\href {https://doi.org/10.1364/OE.26.023265} {\bibfield  {journal} {\bibinfo  {journal} {Optics Express}\ }\textbf {\bibinfo {volume} {26}},\ \bibinfo {pages} {23265} (\bibinfo {year} {2018})}\BibitemShut {NoStop}%
\bibitem [{\citenamefont {Kopp}\ \emph {et~al.}(1998)\citenamefont {Kopp}, \citenamefont {Fan}, \citenamefont {Vithana},\ and\ \citenamefont {Genack}}]{Kopp1998}%
  \BibitemOpen
  \bibfield  {author} {\bibinfo {author} {\bibfnamefont {V.~I.}\ \bibnamefont {Kopp}}, \bibinfo {author} {\bibfnamefont {B.}~\bibnamefont {Fan}}, \bibinfo {author} {\bibfnamefont {H.~K.~M.}\ \bibnamefont {Vithana}},\ and\ \bibinfo {author} {\bibfnamefont {A.~Z.}\ \bibnamefont {Genack}},\ }\bibfield  {title} {\bibinfo {title} {Low-threshold lasing at the edge of a photonic stop band in cholesteric liquid crystals},\ }\href {https://doi.org/10.1364/ol.23.001707} {\bibfield  {journal} {\bibinfo  {journal} {Optics Letters}\ }\textbf {\bibinfo {volume} {23}},\ \bibinfo {pages} {1707} (\bibinfo {year} {1998})}\BibitemShut {NoStop}%
\bibitem [{\citenamefont {Sala}\ and\ \citenamefont {Karpierz}(2012)}]{sala2012modeling}%
  \BibitemOpen
  \bibfield  {author} {\bibinfo {author} {\bibfnamefont {F.~A.}\ \bibnamefont {Sala}}\ and\ \bibinfo {author} {\bibfnamefont {M.~A.}\ \bibnamefont {Karpierz}},\ }\bibfield  {title} {\bibinfo {title} {Modeling of molecular reorientation and beam propagation in chiral and non-chiral nematic liquid crystals},\ }\href {https://doi.org/10.1364/OE.20.013923} {\bibfield  {journal} {\bibinfo  {journal} {Optics Express}\ }\textbf {\bibinfo {volume} {20}},\ \bibinfo {pages} {13923} (\bibinfo {year} {2012})}\BibitemShut {NoStop}%
\bibitem [{\citenamefont {Jungling}\ and\ \citenamefont {Chen}(1994)}]{jungling1994study}%
  \BibitemOpen
  \bibfield  {author} {\bibinfo {author} {\bibfnamefont {S.}~\bibnamefont {Jungling}}\ and\ \bibinfo {author} {\bibfnamefont {J.~C.}\ \bibnamefont {Chen}},\ }\bibfield  {title} {\bibinfo {title} {A study and optimization of eigenmode calculations using the imaginary-distance beam-propagation method},\ }\href {https://doi.org/10.1109/3.309869} {\bibfield  {journal} {\bibinfo  {journal} {IEEE Journal of Quantum Electronics}\ }\textbf {\bibinfo {volume} {30}},\ \bibinfo {pages} {2098} (\bibinfo {year} {1994})}\BibitemShut {NoStop}%
\bibitem [{\citenamefont {Dabrowski}()}]{dabrowski1110mixture}%
  \BibitemOpen
  \bibfield  {author} {\bibinfo {author} {\bibfnamefont {R.}~\bibnamefont {Dabrowski}},\ }\href@noop {} {\bibinfo {title} {1110 nematic mixture}},\ \bibinfo {howpublished} {Personal communication / Sample source information},\ \bibinfo {note} {synthesized at the Military University of Technology, Warsaw}\BibitemShut {NoStop}%
\bibitem [{\citenamefont {Piccardi}\ \emph {et~al.}(2013)\citenamefont {Piccardi}, \citenamefont {Alberucci},\ and\ \citenamefont {Assanto}}]{Piccardi2013IJMS}%
  \BibitemOpen
  \bibfield  {author} {\bibinfo {author} {\bibfnamefont {A.}~\bibnamefont {Piccardi}}, \bibinfo {author} {\bibfnamefont {A.}~\bibnamefont {Alberucci}},\ and\ \bibinfo {author} {\bibfnamefont {G.}~\bibnamefont {Assanto}},\ }\bibfield  {title} {\bibinfo {title} {Nematicons and their electro-optic control: Light localization and signal readdressing via reorientation in liquid crystals},\ }\href {https://doi.org/10.3390/ijms141019932} {\bibfield  {journal} {\bibinfo  {journal} {International Journal of Molecular Sciences}\ }\textbf {\bibinfo {volume} {14}},\ \bibinfo {pages} {19932} (\bibinfo {year} {2013})}\BibitemShut {NoStop}%
\end{thebibliography}

%

\end{document}